\documentclass[twocolumn]{aastex62} 

\usepackage{CLASSY_Preamble}

\newcommand{\ciii}{\ion{C}{3}]}

\newcommand{\civ}{\ion{C}{4}}

\newcommand{\heii}{\ion{He}{2}}

\newcommand{\oiii}{[\ion{O}{3}]}
\newcommand{\oiiiuv}{\ion{O}{3}]}
\newcommand{\feiii}{[\ion{Fe}{3}]}
\newcommand{\feii}{[\ion{Fe}{2}]}
\newcommand{\hii}{\ion{H}{2}}

\newcommand{\ha}{H$\alpha$}
\newcommand{\hb}{H$\beta$}
\newcommand{\hg}{H$\gamma$}
\newcommand{\hd}{H$\delta$}
\newcommand{\siii}{[\ion{S}{3}]}

\newcommand{\nii}{[\ion{N}{2}]}
\newcommand{\niii}{[\ion{N}{3}]}
\newcommand{\niv}{\ion{N}{4}]}
\newcommand{\sii}{[\ion{S}{2}]}

\newcommand{\oi}{[\ion{O}{1}]}

\newcommand{\ariv}{[\ion{Ar}{4}]}
\newcommand{\ariii}{[\ion{Ar}{3}]}
\newcommand{\neiii}{[\ion{Ne}{3}]}

\newcommand{\arii}{[\ion{Ar}{2}]}
\newcommand{\nevi}{[\ion{Ne}{6}]}
\newcommand{\oiv}{[\ion{O}{4}]}
\newcommand{\ov}{[\ion{O}{5}]}
\newcommand{\nev}{[\ion{Ne}{5}]}
\newcommand{\neii}{[\ion{Ne}{2}]}
\newcommand{\siv}{[\ion{S}{4}]}
\newcommand{\neiv}{[\ion{Ne}{4}]}

\newcommand{\hhuma}{Hu$\alpha$}
\newcommand{\hpfa}{Pf$\alpha$}
\newcommand{\mgvii}{[\ion{Mg}{7}]}
\newcommand{\mgv}{[\ion{Mg}{5}]}

\usepackage{lipsum}
\usepackage{multirow}
\usepackage{gensymb}

\graphicspath{{./}{figures/}}

\received{11-Feb-2025}
\revised{26-Mar-2025}
\accepted{07-Apr-2025} 
\submitjournal{ApJ}

\shortauthors{Mingozzi et al.}
\shorttitle{Exploring the mysterious high-ionization source powering [Ne~V] in SBS0335-052~E}

\begin{document}

\title{Exploring the mysterious high-ionization source powering [Ne~V] in high-$z$ analog SBS0335-052~E with JWST/MIRI}

\author[0000-0003-2589-762X]{Matilde Mingozzi}
\affiliation{AURA for ESA, Space Telescope Science Institute, 3700 San Martin Drive, Baltimore, MD 21218, USA}

\author[0000-0002-0191-4897]{Macarena Garcia del Valle-Espinosa}
\affiliation{Space Telescope Science Institute, 3700 San Martin Drive, Baltimore, MD 21218, USA}

\author[0000-0003-4372-2006]{Bethan L. James}
\affiliation{AURA for ESA, Space Telescope Science Institute, 3700 San Martin Drive, Baltimore, MD 21218, USA}

\author[0000-0001-9719-4080]{Ryan J. Rickards Vaught}
\affiliation{Space Telescope Science Institute, 3700 San Martin Drive, Baltimore, MD 21218, USA}

\author[0000-0001-8587-218X]{Matthew Hayes}
\affiliation{Stockholm University, Department of Astronomy and Oskar Klein Centre for Cosmoparticle Physics, AlbaNova University Centre, SE-10691, Stockholm, Sweden}

\author[0000-0001-5758-1000]{Ricardo O. Amor\'{i}n}
\affiliation{Instituto de Astrof\'{i}sica de Andaluc\'{i}a (CSIC), Apartado 3004, 18080 Granada, Spain}

\author[0000-0003-2685-4488]{Claus Leitherer}
\affiliation{Space Telescope Science Institute, 3700 San Martin Drive, Baltimore, MD 21218, USA}

\author[0000-0003-4137-882X]{Alessandra Aloisi}
\affiliation{Space Telescope Science Institute, 3700 San Martin Drive, Baltimore, MD 21218, USA}
\affiliation{Astrophysics Division, Science Mission Directorate, NASA Headquarters, 300 E Street SW, Washington, DC 20546, USA}

\author[0000-0001-9162-2371]{Leslie Hunt}
\affiliation{INAF - Osservatorio Astrofisico di Arcetri, Largo E. Fermi 5, I-50125, Firenze, Italy}

\author[0000-0002-9402-186X]{David Law}
\affiliation{Space Telescope Science Institute, 3700 San Martin Drive, Baltimore, MD 21218, USA}

\author[0000-0002-3703-0719]{Chris T. Richardson}
\affiliation{Elon University, 100 Campus Drive, Elon, NC 27278, USA}

\author[0000-0002-1655-6041]{Aidan Pidgeon}
\affiliation{Space Telescope Science Institute, 3700 San Martin Drive, Baltimore, MD 21218, USA}

\author[0000-0002-2644-3518]{Karla Z. Arellano-C\'{o}rdova}
\affiliation{Institute for Astronomy, University of Edinburgh, Royal Observatory, Edinburgh EH9 3HJ, UK}

\author[0000-0002-4153-053X]{Danielle A. Berg}
\affiliation{Department of Astronomy, The University of Texas at Austin, 2515 Speedway, Stop C1400, Austin, TX 78712, USA}

\author[0000-0002-0302-2577]{John Chisholm}
\affiliation{Department of Astronomy, The University of Texas at Austin, 2515 Speedway, Stop C1400, Austin, TX 78712, USA}

\author[0000-0003-4857-8699]{Svea Hernandez}
\affiliation{AURA for ESA, Space Telescope Science Institute, 3700 San Martin Drive, Baltimore, MD 21218, USA}

\author[0000-0002-1706-7370]{Logan Jones}
\affiliation{Space Telescope Science Institute, 3700 San Martin Drive, Baltimore, MD 21218, USA}

\author[0000-0002-5320-2568]{Nimisha Kumari}
\affiliation{AURA for ESA, Space Telescope Science Institute, 3700 San Martin Drive, Baltimore, MD 21218, USA}

\author[0000-0001-9189-7818]{Crystal L. Martin}
\affiliation{Department of Physics, University of California, Santa Barbara, Santa Barbara, CA 93106, USA}

\author[0000-0002-5269-6527]{Swara Ravindranath}
\affiliation{Astrophysics Science Division, NASA Goddard Space Flight Center, 8800 Greenbelt Road, Greenbelt, MD 20771, USA}
\affiliation{Center for Research and Exploration in Space Science and Technology II, Department of Physics, Catholic University of America, 620 Michigan Ave N.E., Washington DC 20064, USA}

\author[0000-0002-3258-3672]{Livia Vallini}
\affiliation{INAF – Osservatorio di Astrofisica e Scienza dello Spazio di Bologna, Via Piero Gobetti, 93/3, I-40129 Bologna, Italy}

\author[0000-0002-9217-7051]{Xinfeng Xu}
\affiliation{Department of Physics and Astronomy, Northwestern University, 2145 Sheridan Road, Evanston, IL 60208, USA}
\affiliation{Center for Interdisciplinary Exploration and Research in Astrophysics (CIERA), Northwestern University, 1800 Sherman Avenue, Evanston, IL 60201, USA}


\correspondingauthor{Matilde Mingozzi} 
\email{mmingozzi@stsci.edu}



\begin{abstract}
Nearby blue compact dwarf galaxies (BCDs) are considered analogs to objects from the Epoch of Reionization revealed by JWST, having similar low stellar mass, low metallicity, and high specific star-formation rate. 
Thus, they represent ideal local laboratories detailed for multi-wavelength studies of their properties and mechanisms that shape them.
We report the first JWST MIRI/MRS observations of the BCD SBS~0335-052~E, analyzing MIR emission lines tracing different levels of ionization (e.g., \neii, \siv, \neiii, \oiv, \nev) of the ionized gas.
SBS~0335-052~E MIR emission is characterized by a bright point source, located in one of the youngest and most embedded stellar clusters ($t\sim3$~Myr, $A_V\sim15$), and underlying extended high-ionization emission (i.e., \oiv, \nev) from the surroundings of the older and less dusty stellar clusters ($t< 20 $~Myr, $A_V\sim8$). 
From the comparison with state-of-the-art models, we can exclude shocks, X-ray binaries, and old stellar populations as the main sources of ionization. 
Interestingly, a 4-8\% contribution of a $\sim10^5$~M$_\odot$ intermediate massive black hole (IMBH) is needed to justify the strong \nev/\neii\ and would be consistent with optical/UV line ratios from previous studies. However, even IMBH models cannot explain the strongest \oiv/\neiii. 
Also, star-forming models (regardless of including X-ray binaries) struggle to reproduce even the lower ionization line ratios (e.g., \siv/\neii) typically observed in BCDs. 
Overall, while current models suggest the need to account for an accreting IMBH in this high-$z$ analog, limitations still exist in predicting high-ionization emission lines (I.P.~$>54$~eV) when modeling these low-metallicity environments, thus other sources of ionization cannot be fully ruled out.
\end{abstract} 


\section{Introduction} \label{sec:intro}
\looseness=-2
In the JWST era, surveys such as the {\it JWST Advanced Deep Survey} (JADES), the {\it Cosmic Evolution Early Release Science Survey} (CEERS) and the {\it Ultradeep NIRSpec and NIRCam Observations before the Epoch of Reionization} (UNCOVER) have started to revolutionize our knowledge of the early Universe up to $z\sim14$ (e.g., \citealt{bagley2023,bunker2023b,bezanson2024,carniani2024}), exploring the epoch of reionization (EoR, $z>6$) as never done before.
Several studies over the past decade have suggested that low-mass, metal-poor and highly star-forming galaxies can play a crucial role in the EoR (e.g., \citealt{wise2014,robertson2015,stark2015}), as now supported by recent JWST works probing the details of ultra-faint galaxies (e.g., \citealt{atek2024,simmonds2024}). 
JWST has also started to reveal an unprecedented number of active galactic nuclei (AGN) in the first billion years, comprising both intrinsically faint and reddened sources (e.g., \citealt{harikane2023,maiolino2023b,labbe2023,kocevski2024,ubler2024,juodzbalis24}), including a new family of objects, the so-called Little Red Dots  \citep[e.g.,][]{furtak2023,kokorev2024,matthee2024}. 
The role that these AGN can play in the EoR is currently under investigation \citep[e.g.,][]{dayal2024}, but it is still challenging to unambiguously identify AGN activity in EoR systems.

To detect the presence of faint AGN at $4<z<12$, many works used the detection of broadened Balmer lines (FWHM~$\sim1000-5000$~km/s; e.g., \citealt{harikane2023,maiolino2023b,larson2023,kokorev2023,greene2024,greene2024b}). 
However, the broad component under forbidden lines can simply be too faint to be detected \citep[e.g.,][]{carniani2024a}, leading to possible mis-classification of a pure star-forming system with a fast gas outflow as a broad line AGN. 
AGN classification has been also proposed via the presence of high-ionization rest-UV and blue optical lines (e.g., \niv~$\lambda\lambda1483,7$, \niii~$\lambda1750$, \civ~$\lambda\lambda$1548,51, \neiv~$\lambda$2423, \heii~$\lambda$4686; \citealt{maiolino2023a,kokorev2024,greene2024,backhaus2024}). 
However, optical and UV diagnostics can be less effective in metal-poor systems (e.g., \citealt{feltre2016,ubler2023}), because a hard stellar radiation field and AGN activity can behave similarly in low-metallicity dense environments (e.g., \citealt{leitherer1999,eldridge2017,telford2023}). Thus, metal-poor star-forming systems can show UV and optical high-ionization lines similar to those observed in high-z AGN candidates (e.g., the star-forming RXCJ2248-ID at $z\sim6$, \citealt{topping2024}), making the two scenarios difficult to distinguish (e.g., \citealt{castellano2024,alvarez2024}).
Another proposed method to identify AGN is through photometric variability resulting from changes in AGN mass accretion rate \citep{hayes2024}, but larger samples of AGN at high-$z$ are needed to reduce the statistical and systematic uncertainties. 

In this context, nearby galaxies can provide a precious opportunity for us to understand the characteristics of the interstellar medium (ISM) of these primeval objects and how to distinguish between star-formation and AGN activity within them, given the high signal-to-noise (S/N) of their spectra and the possibility to perform spatially resolved and multi-wavelength studies. 
In particular, nearby blue compact dwarf galaxies (BCDs) are ideal analogs to these high-$z$ systems, as they share similar properties in terms of stellar mass, intense star formation activity, and poor metal content, as also suggested by recent JWST studies \citep[e.g.,][]{schaerer2022,brinchmann2023}.
Many studies have explored these objects to understand how to constrain their ISM properties using their optical and UV emission (e.g., \citealt{james2009,izotov2012,berg2019,berg2021,senchyna2021a,senchyna2023,mingozzi2022,mingozzi2024,kumari2024}).
In this context, the mid-infrared (MIR) wavelength range covers a large range of ionization potentials for the same species (e.g., 20-126~eV, \neii~$\lambda$21.56$\mu$m, \neiii~$\lambda$15.56$\mu$m, \nev~$\lambda\lambda$14.32,24.3$\mu$m, \nevi~$\lambda$7.65$~\mu$m) at lower excitation energies. 
While these lines are not accessible in the high-$z$ universe, they can provide an important constraint on the gas ionization mechanisms and, by also probing other phases (i.e., warm molecular gas with H$_2$ rotational lines; dust with polycyclic aromatic hydrocarbons, PAHs), they can give a broad overview of the ISM conditions.

In this work, we focus on the galaxy SBS~0335-052~E ($z\sim 0.01352$; 1"~$\sim280$~pc), discovered by \citet{izotov1990}. This object is a nearby BCD, extremely metal-poor ($\sim 5\%$~Z$_\odot$; 12+log(O/H)~$\sim7.2-7.3$; \citealt{papaderos2006}) and characterized by a compact vigorous starburst (specific star formation rate, log($sSFR$/Gyr$^{-1}$)~$ \sim -8.13$; \citealt{remyruyer2015}) and massive young clusters (\citealt{papaderos1998,reines2008,adamo2010} and references therein). 
This galaxy is also well-known for complex kinematics and a large scale \ha\ outflow \citep{herenz2023} and high-ionization optical and UV emission lines typically observed in the EoR (e.g., \heii~$\lambda$4686, \civ~$\lambda\lambda1548,51$, \heii~$\lambda$1640; \citealt{herenz2017,kehrig2018,wofford2021}). 
In particular, its extended nebular \heii~$\lambda$4686 emission requires energy beyond 54.4~eV (i.e., $>4$~Ryd, $\lambda<228$~\AA), indicative of a very hard ionizing spectral energy distribution (SED), that only rotating metal-free or extremely metal-poor massive stars seem capable of reproducing (\citealt{kehrig2018}; see also \citealt{wofford2021}). 
Recent studies have also suggested that SBS~0335-052~E may host an intermediate massive black hole (IMBH) of $\sim10^3-10^5$~M$_\odot$ to explain its \nev~$\lambda$3426 emission (\citealt{hatano2024}, see also \citealt{thuan2005}), requiring energies beyond 97~eV (i.e., $>7.1$~Ryd, $\lambda<128$~\AA), as well as possible near IR variability \citep{hatano2023}. 
IMBHs (see \citealt{greene2020} for a review) could represent the missing piece between stellar and supermassive BHs and would help explain the rapid accretion processes in place at high-$z$. 
Thus, it is important to understand if galaxies such as SBS~0335-052~E can host an accreting IMBH, and, if so, which are the best diagnostics to trace its activity and impact on the ISM.

To this end, we exploit JWST/MIRI Medium Resolution Spectroscopy (MRS) integral field spectroscopy (IFS) data to investigate the possible accreting IMBH or other non-stellar source ionizing the high-$z$ analog SBS~0335-052~E's ISM, using spatially resolved high ionization MIR lines.
Indeed, SBS 0335-052~E intense SF activity makes this galaxy particularly bright in the MIR, which represents $\sim75$\% of its total luminosity \citep{plante2002}. 
Spitzer IRS $5.3-35$~$\mu$m spectrum already revealed strong MIR emission lines (i.e., \siv~$\lambda$10.51, \neiii~$\lambda$15.55, \oiv~$\lambda$25.89), warm molecular gas features (i.e., H$_2$~$0-0$~S(3)) and the absence of PAHs \citep{houck2004,hao2009}. 
Nevertheless, MIRI/MRS offers the considerable advantage of spatially resolving the MIR emission (down to $\sim 0.4-0.9$", $\sim 112-252$~pc) and thus revealing the location of very high ionization extended emission (i.e., \nev~$\lambda$14.32$\mu$m) for the first time in BCDs. 

In Sec.~\ref{sec:observations}, we present the observations and data reduction. 
In Sec.~\ref{sec:data-analysis}, we show the methods we adopted for our data analysis, including spectral fitting, and literature photoionization and shock models that we used to interpret our results, reported in Sec.~\ref{sec:results}. 
In Sec.~\ref{sec:discussion}, we discuss our findings as well as possible limitations in current models, while in Sec.~\ref{sec:conclusion}, we conclude with a summary of our work. 
Throughout this work we assume a flat $\Lambda$CDM cosmology (H$_0 = 70$~km/s/Mpc, $\Omega=0.3$) and 12+log(O/H)$_\odot = 8.69$ \citep{asplund2009}.

\section{Observations and Data Reduction}\label{sec:observations}
\begin{figure*}
\begin{center}
    \includegraphics[width=0.9\textwidth]{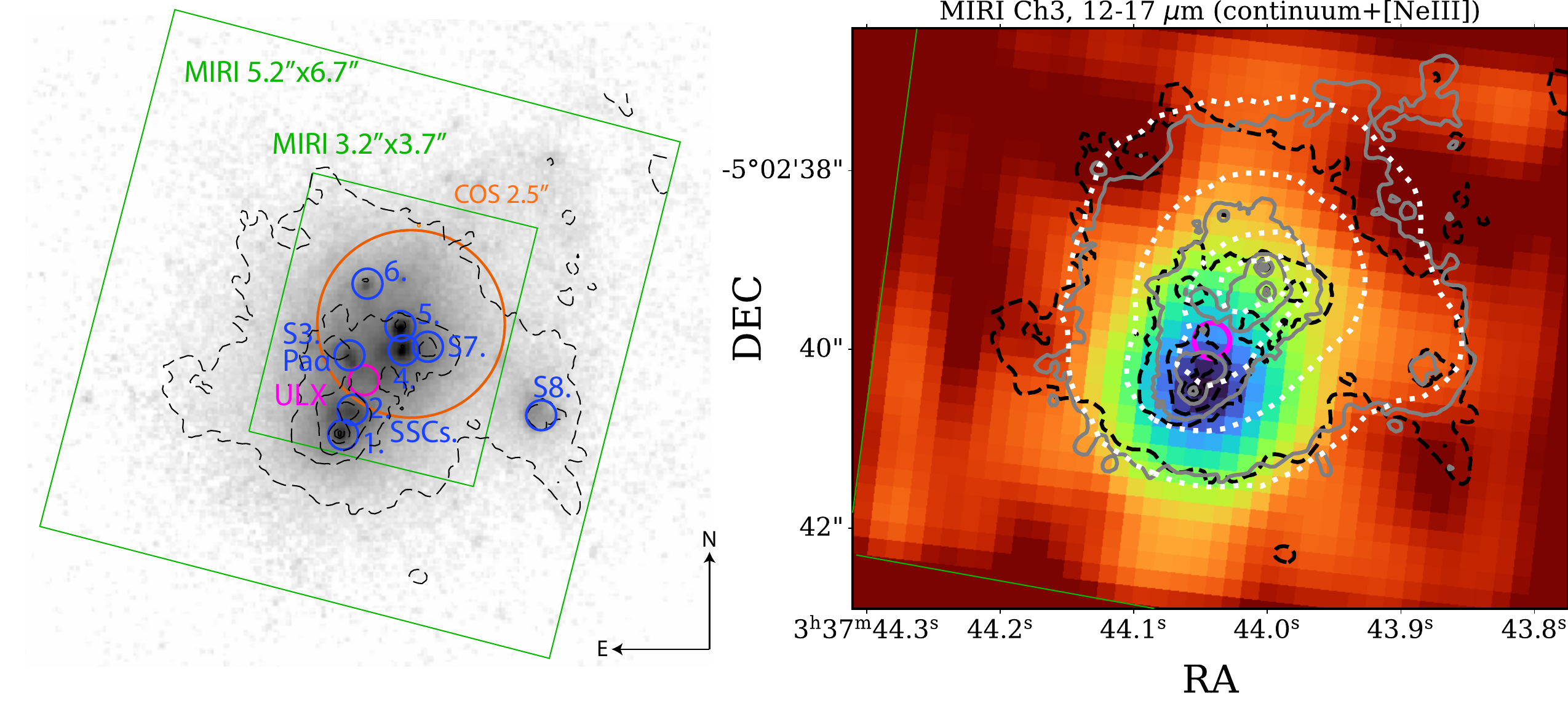}
\end{center}
\caption{Left panel: Overview on SBS~0335-052~E (1"~$\sim280$~pc). HST ACS F150LP image (PID: 16209; PI Hayes) with superimposed MIRI/MRS Channel 1 and 3 FOVs in green, and the HST ACS WFC FR656N (PID: 10575, PI Oestlin) contours in black; the six SSCs positions (SSCs1,2,3,4,5,6, $t\sim3-13$~Myr; see Tab.~\ref{tab:compilationResults}) and the three further young star-forming regions (S7, S8 and S3-Pa$\alpha$) found by \citet{thompson2009} are indicated in blue; the magenta circle with a diameter equal to Chandra’s positional uncertainty indicates the ULX position \citep[][their Tab.~3]{prestwich2013}; the COS 2.5" aperture indicated in orange covers the brightest UV emitting region analyzed in the CLASSY survey \citep{berg2022,james2022}. North is up, east is to the left. Right panel: 12-17~$\mu$m slice from MIRI/MRS Channel 3. The log-scale color map allows highlighting of the MIRI/MRS PSF pattern due to a bright point source located close to the position of SSC1, as well as extended emission coming from the other stellar clusters. The black and gray contours show the H$\alpha$ and UV emission from the HST ACS WFC FR656N and F150LP, respectively. The white contours indicate the \heii~$\lambda$4684 emission from MUSE data (ID~096.B-0690A; PI Hayes).}
\label{fig:sbs-f140lp}
\end{figure*}
SBS 0335-052~E data were acquired as part of the Cycle 2 JWST program PID 4278 (PI: Mingozzi) on February 11, 2024. 
In particular, the galaxy was observed with JWST/MIRI MRS using the three MRS grating settings (SHORT, MEDIUM, and LONG; $\lambda \sim 5-28$~$\mu m$), and the FAST readout mode. Just one pointing was used since MIRI/MRS covers the majority of SBS~0335-052~E's active starburst activity with Channel~1 (3.2"$\times$3.7") and fully samples it in Channel~4 (6.6"$\times$7.7"). Given the JWST pointing accuracy of $\sim0.1$", target acquisition was not required. A 4-point dither pattern was also applied to achieve optimal sampling throughout the MRS field of view (FOV) and to identify and remove detector artefacts. Each pointing was observed with 50 groups per integration for a total on-source exposure time of 1665~s. 
Dedicated background observations were performed with half the integrations (same group number but half the dithering) to accurately measure and correct for the thermal background. 
Background and science observations were connected via a fixed offset and placed in an uninterruptible sequence, and a PA orientation range was required to obtain simultaneous imaging observations using the F770W, F1130W, and F1800W filters, which will be presented in a future publication. 

We retrieved MIRI/MRS raw products from Mikulski Archive for Space Telescopes (MAST), and we processed the data with v1.16.0 of the JWST calibration pipeline \citep{jwstpipeline}, following the \href{https://github.com/spacetelescope/jwst-pipeline-notebooks/}{jupyter notebook} provided by the MIRI team. As extra steps to the standard version of the MIRI pipeline, following the recommendations from the JWST pipeline reduction team, we masked additional regions affected by cosmic showers at the end of Stage~1 (\href{https://github.com/STScI-MIRI/ShowerMasking/blob/main/manual_shower_masking.ipynb }{ShowerMasking Notebook}) and we applied residual fringe corrections in Stage~2. 
We subtracted the background in Stage~3, using the master background created by the pipeline.
In Stage~3, we created the final datacubes per band, to optimize the fringing correction, obtaining 12 datacubes centered to the same coordinates and keeping the native pixel scale size (i.e., 0.13", 0.17", 0.20" and 0.35", for Channel 1, 2, 3 and 4, respectively). Nevertheless, we note that the spaxels surrounding the bright point source that we further discuss below, within a circular region with a radius equal to the MIRI/MRS PSF FWHM, are still significantly affected by fringing, especially in Channels~3 and 4.
The FOV progressively increases from 3.2"$\times$3.7" in Channel 1, to 6.6"$\times$7.7" in Channel 4, as well as the MIRI/MRS PSF and spatial resolution, from $\sim 0.4"$ in Channel 1 to $\sim 0.9"$ in Channel 4.

In Fig.~\ref{fig:sbs-f140lp}, we show the position of MIRI/MRS FOV superimposed on the HST ACS F150LP UV emission (left panel) and the 12-17~$\mu$m slice from MIRI/MRS Channel 3 (right panel).  
The UV emission (map on the left panel, gray contours on the right panel) clearly shows the six young, massive and compact ``super star clusters'' (SSC1, SSC2, SSC3, SSC4, SSC5, SSC6) revealed in this galaxy \citep{thuan1997}. SSCs1-6 clusters have ages and masses in the range $\sim3-15$~Myr and $\sim3-30 \times 10^5$~M$_\odot$ (\citealt{reines2008,adamo2010}; see also Tab.~\ref{tab:compilationResults}), and luminosity of $\sim10^8$~L$_\odot$, all within $\sim2$"~$\sim500$~pc \citep{thuan1997}. 
One of them, SSC1, is also known for its enhanced obscuration \citep{hunt2001}.
The \heii~$\lambda$4686 high-ionization emission revealed with MUSE (white dotted contour on the right panel; seeing $\sim0.7-0.8$") follows the UV emission and peaks around SSCs4,5.
The HST H$\alpha$ black dashed contours highlight the three additional actively young star-forming regions, S3-Pa$\alpha$, S7 and S8, discovered by \citet{thompson2009} by analyzing SBS~0335-052~E Pa$\alpha$ emission. 
The MIRI/MRS image (right panel) reveals the clear MIRI/MRS PSF pattern due to a very bright point source located at the position of SSC1 as well as extended emission covering the other stellar clusters, as we will discuss in Sec.~\ref{sec:data-analysis}. 

\section{Data Analysis and Methods}\label{sec:data-analysis}
By a first visual inspection of the Stage~3 products, we noticed a point-source like emission dominating the continuum and emission-line fluxes close to the position of SSC1 (see Fig.~\ref{fig:sbs-f140lp} right panel). 
To spatially investigate the source of emission (Sec.~\ref{sec:results} and Sec.~\ref{sec:discussion}), it was essential to determine the spatial extent of the underlying extended emission. This was achieved by accurately modeling and isolating the point source contribution, ending up with a point-source subtracted dataset and a point-source spectrum.
The details of this subtraction procedure are described in App.~\ref{app:psf-sub}. 
In Sec.~\ref{sec:fitting} and Sec.~\ref{sec:dust}, we describe the data analysis applied to the point source and extended emission spectra.

\subsection{Emission-line fitting}\label{sec:fitting}
\begin{figure*}[t]
\begin{center}
    \includegraphics[width=0.75\textwidth]{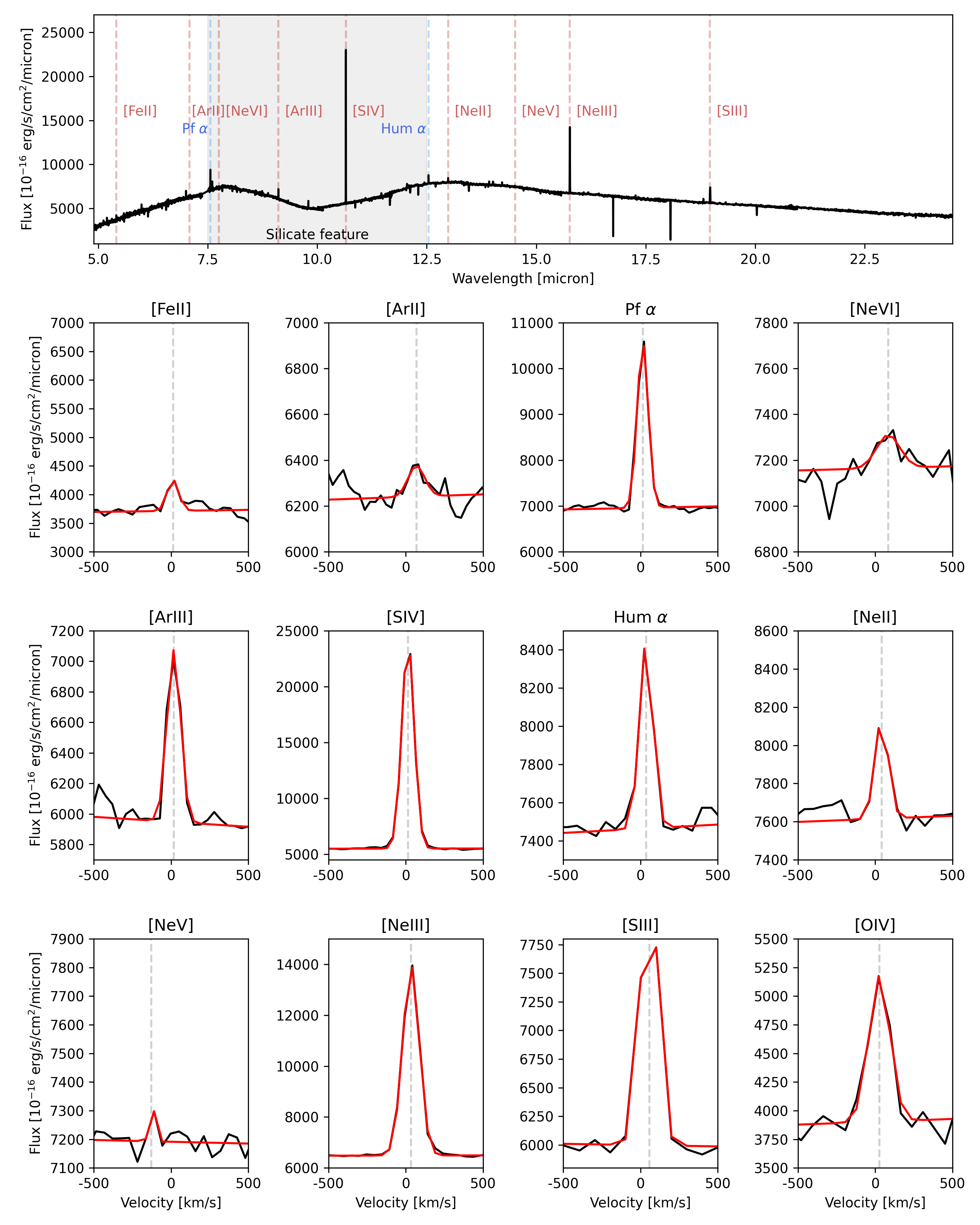}
\end{center}
\caption{Point-source (SSC1) spectrum with the set of emission lines highlighted in blue for the hydrogen-recombination lines and red for the metallic-forbidden lines. The bottom panels display a zoom-in around each emission line considered in this work, with the best-fit model shown in red. The point-source spectrum also shows a very high attenuation, with $A_V\sim 20$. The larger contribution of the dust can be appreciated by the silicate feature at 9.7 $\mu$m (gray shaded area), most pronounced in the point source spectrum. \nevi~$\lambda$~7.65, \nev~$\lambda$~14.32 and \arii~$\lambda$~6.98 have $S/N\lesssim2.5$, so they cannot be considered detections. All the other lines have $S/N>3$.}
\label{fig:1dspecpsf}
\end{figure*}
\begin{figure*}[t]
\begin{center}
    \includegraphics[width=0.75\textwidth]{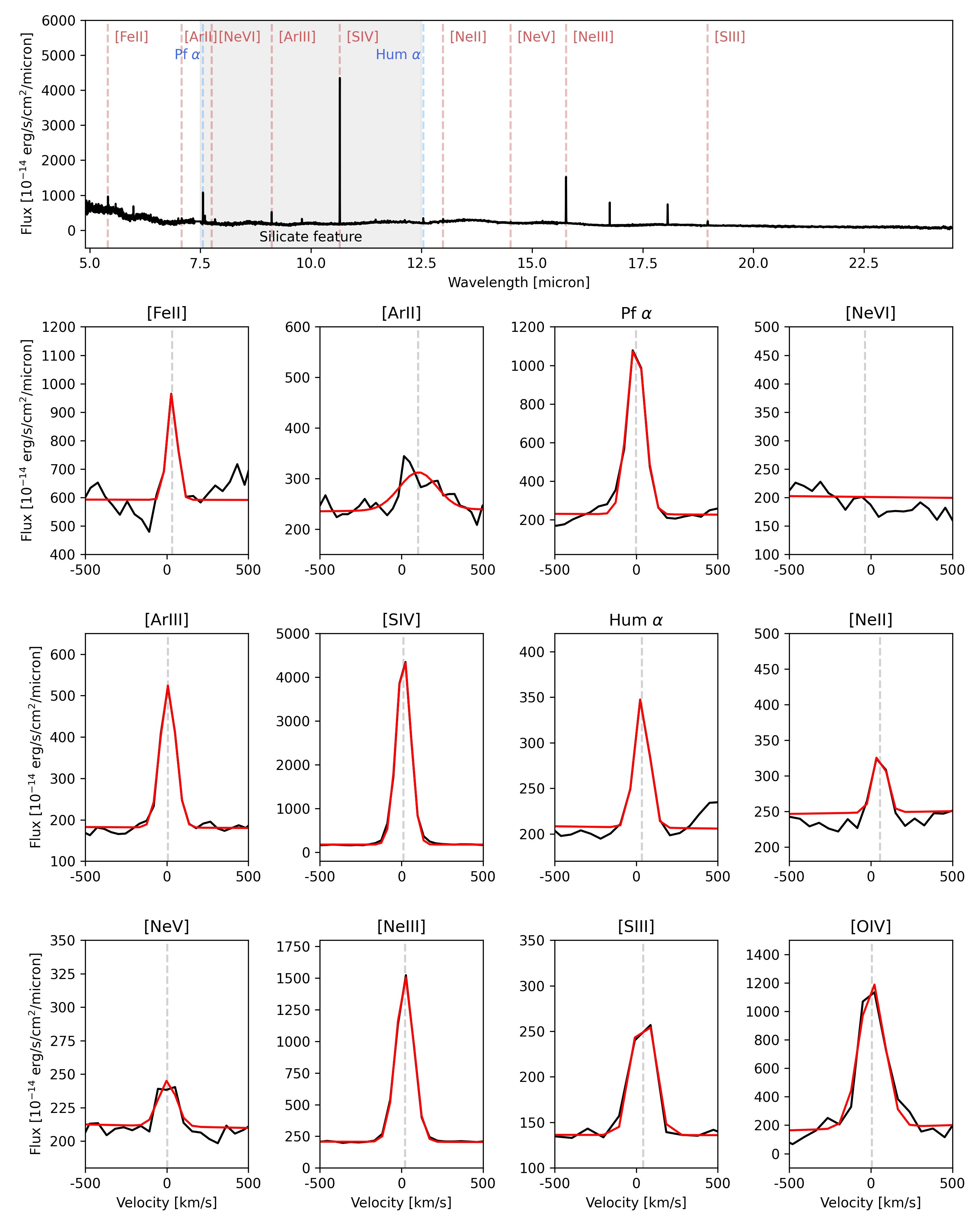}
\end{center}
\caption{Same as Fig.~\ref{fig:1dspecpsf} for the spectrum extracted at the ULX position (see Fig.~\ref{fig:sbs-f140lp}), ULX spectrum is extracted from a region with variable diameter depending on the channel, matching the FWHM of the PSF (i.e., $FWHM_{Ch1,2,3,4} =$~0.4", 0.5", 0.6", 0.9" \citep{law2023}). \nevi~$\lambda$~7.65 and \arii~$\lambda$~6.98 are not detected, while all the other marked lines have $S/N>3$.}
\label{fig:1dspeculx}
\end{figure*}
\begin{figure*}[t]
\begin{center}
    \includegraphics[width=0.75\textwidth]{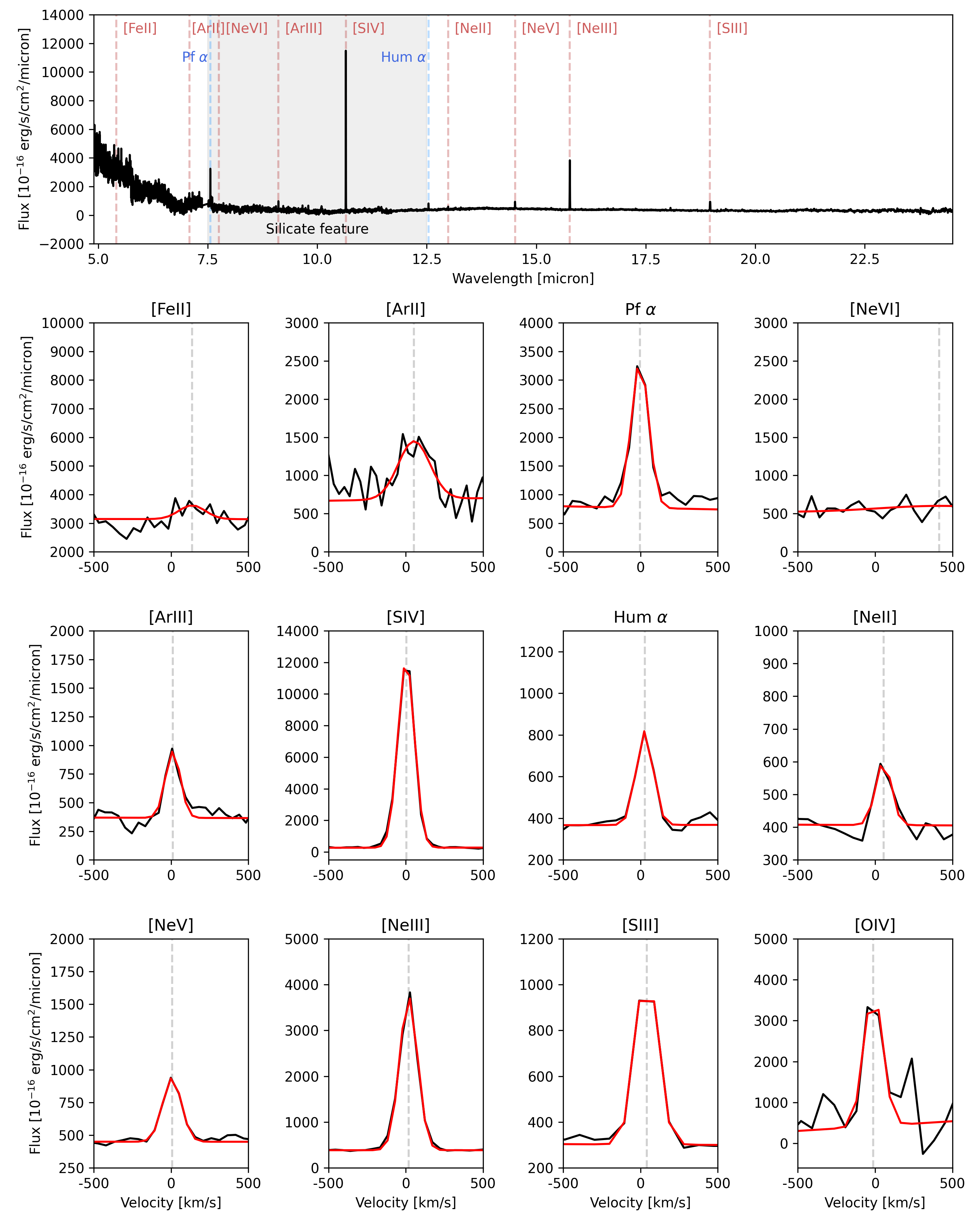}
\end{center}
\caption{Same as Fig.~\ref{fig:1dspecpsf} for the region covering the peak of the UV emission and \heii~$\lambda$4686 emission observed with HST and MUSE, respectively (see Fig.~\ref{fig:sbs-f140lp}). As shown in Fig.~\ref{fig:sbs-f140lp}, this region also covers SSCs~3,4,5,6, S3-Pa$\alpha$, S7, and the ULX. \feii~$\lambda$5.34 and \nevi~$\lambda$~7.65 are not detected, while all the other lines have $S/N>3$. }
\label{fig:1dspecnev}
\end{figure*}
In each spectrum, we fitted several MIR emission lines, performing the line fitting locally around each line in a $\pm 1000$ km/s window centered at the observed wavelength of the line of interest. 
In particular, we used a combination of a Gaussian profile and a one-degree polynomial to account for emission-line and surrounding continuum emission in each spaxel of the point-source subtracted band datacube slices, as well as in the 1-D spectrum of the bright point-source and the spectrum extracted at the ULX position (shown in Fig.~\ref{fig:sbs-f140lp}). 
The 1-D spectrum of the bright point source is obtained by the best-fit model of the PSF photometry analysis described in App.~\ref{app:psf-sub}, while the ULX spectrum is extracted from a region with variable diameter depending on the channel, matching the FWHM of the PSF (i.e., $FWHM_{Ch1,2,3,4} =$~0.4", 0.5", 0.6", 0.9"; \citealt{law2023}). 
Tab.~\ref{tab:listEmissionLines} compiles the list of emission lines analyzed, together with their rest-frame wavelengths and ionization potentials (I.P.).
\begin{table}[]
    \centering
       \caption{List of the fitted MIR emission lines, including their rest-wavelength and ionization potential (I.P.). }
    \begin{tabular}{lcc} 
    \hline\hline
        Line & $\lambda$ & I.P.  \\
        & ($\mu$m) & (eV) \\ 
        \hline
        \feii~  & 5.34  & 7.90  \\
        \arii~  & 6.98  & 15.76  \\
        \hpfa~  & 7.46  & 13.6   \\
        \nevi~  & 7.65  & 126.25 \\
        \ariii~ & 8.99  & 27.63  \\
        \siv~   & 10.51 & 34.79  \\
        \hhuma~ & 12.37 & 13.6   \\
        \neii~  & 12.81 & 21.56  \\
        \nev~   & 14.32 & 97.12  \\
        \neiii~ & 15.56 & 40.96  \\
        \siii~  & 18.71 & 23.34  \\
        \nev~   & 24.32 & 97.12  \\
        \oiv~   & 25.89 & 54.95  \\
    \end{tabular}
    \label{tab:listEmissionLines}
\end{table}

Fig.~\ref{fig:1dspecpsf}, \ref{fig:1dspeculx} and \ref{fig:1dspecnev} show the point source spectrum (SSC1), the spectrum extracted from the spaxels covering the ULX position, and a spectrum from the north-west region covering the older clusters SSCs3,4,5,6 and S7, the peak of UV emission and the ULX (see Fig.~\ref{fig:sbs-f140lp}), respectively, as well as the best-fit (in red) for the MIR emission lines. 
In the following, we consider line detections as those with fluxes with signal-to-noise (S/N) higher than 3 and upper limits for measurements with S/N$\gtrsim$2. 
In particular, we note that emission lines typically observed in AGN (e.g., \citealt{dasyra2024}), such as \nevi~$\lambda$~7.65 (tentatively fitted only in the bright point source spectrum but with $S/N\lesssim2.5$), \mgv~$\lambda$~5.61 (I.P.~$=109.3$~eV) and \mgvii~$\lambda$~5.5 (I.P.~$=186.76$~eV) are not detected. 

\subsection{Dust attenuation}\label{sec:dust}
The emission-line fluxes were converted from the original MJy/sr into erg/s/cm$^2$/arcsec$^2$ and corrected for the total foreground Galactic reddening along the line of sight using the $E(B-V)$ values determined for this object by \citet{berg2022} and the \citet{gordon2023} dust extinction relationship (see also \citealt{gordon2009,gordon2021,decleir2022,fitzpatrick2019}). 

When exploring the spectrum of the point source SSC1, we noticed a wide absorption in the 7.5-12.5~$\mu$m range, corresponding to the silicate feature at 9.7~$\mu$m (highlighted in gray in Fig.~\ref{fig:1dspecpsf}, \ref{fig:1dspeculx} and \ref{fig:1dspecnev}). 
This feature (revealed also in SBS~0335-052~E Spitzer spectrum; \citealt{houck2004}) pinpoints the presence of dust within this region, as already reported by \citet{hunt2014}. 
Thus, an intrinsic dust attenuation correction has to be applied to the line fluxes as well. 
Before using the continuum to determine the level of dust attenuation, there are a few extra corrections we needed to implement to the datacubes. 
First, we limited the impact of fringing on the continuum spectra by applying an additional 1D fringing correction using the JWST pipeline tools. 
Also, the data exhibit discontinuities, or ``jumps”, in the flux normalization among the different bands, which we corrected by fixing the normalization to Band A in Channel 1.
With the continuum corrected and optimized for the analysis, we used the Continuum and Feature Extraction software \citep[\texttt{CAFE},][]{marshall2007,cafe} to model the continuum of both the point-source spectrum and the point-source subtracted (extended) continuum in the range $5-20$~$\mu$m from Channel 1, 2 and 3 band datacubes (see App.~\ref{app:psf-sub}). 
\texttt{CAFE} uses a combination of blackbody emission at different temperatures together with a custom attenuation law (compatible with \citealt{gordon2023} law; see also \citealt{gordon2009,gordon2021,decleir2022,fitzpatrick2019}) to describe the overall shape of the continuum at near-to-mid infrared wavelengths. 
For the case of SBS 0335-052~E only two dust components were considered: a cool component, whose temperature falls in the range 40-100 K, and a warm component, with temperature in the range 100-500 K. While both spectra, the point-like source and the extended emission, needed a cool component of $T\sim$80 K, only the point-like source required an additional warm component with $T\sim$150 K to model the continuum shape. 
A more detailed description of the dust attenuation correction will be presented in del Valle-Espinosa et al., in prep.
As a result, the peak of dust attenuation corresponds to the point source SSC1 with $A_V\sim15$, while for the extended region we considered an average value of $A_V\sim8$. We used the derived $A_V$ values and the \citet{gordon2023} dust extinction relationship (see also \citealt{gordon2009,gordon2021,decleir2022,fitzpatrick2019}) to correct for intrinsic dust attenuation.

Although the extinction curve at the MIR wavelengths is almost flat, lines located in the deepest part of the silicate feature (highlighted in gray in Fig.~\ref{fig:1dspecpsf}, \ref{fig:1dspeculx} and \ref{fig:1dspecnev}) can suffer from large attenuation. In the case of SBS~0335-052~E, the dust-corrected fluxes of almost every emission line are increased by a factor of $\sim2$ ($\sim1.2$) in the point-like source (extended) spectrum, except for the \ariii~and \siv~lines — which fall within the silicate feature — whose correction factor is $\sim3$ ($\sim1.6$).
This also implies that only emission-line ratios including \ariii\ and \siv\ lines can be affected by dust attenuation.

\subsection{PSF and pixel size matching}\label{sec:psf-pixel-matching}
In this paper, we want to diagnose the main ionization source by comparing emission-line ratios and models. 
Since MIRI/MRS channels have a wavelength-dependent PSF and a different pixel size, we needed to convolve and rescale our emission-line maps before calculating the ratios of lines in different regions of the MIR spectrum. 
We avoided taking into account the complexity of the PSF profile in each sub-channel and wavelength as we did for the PSF subtraction procedure since it would have added further artefacts and uncertainties. 
Instead, we considered the linear relation between the MIRI/MRS PSF FWHM and $\lambda$ found by \citet[][see their Eq.~1]{law2023}. 
Then, we convolved the line maps with a 2-D Gaussian kernel set to a value equal to the difference in quadrature of the PSF FWHMs at the wavelengths taken into account, similarly to \citet{jones2024}.
Specifically, we created two sets of emission-line maps, convolving them either to the \neiii~$\lambda$~15.56 or the \oiv~$\lambda$~25.89 PSFs ($FWHM_{PSF}\sim 3$~px, 4.5~px, $\sim 0.6$", 0.9", $\sim 168,252$~pc, respectively). 
Finally, we rescaled all the emission maps to Channel~3 pixel size (0.245"), where the majority of the emission lines taken into account in this work are located. 

\subsection{Photoionization and shock models}\label{sec:models}
In order to understand the main source of ionization in Sec.~\ref{sec:results} and Sec.~\ref{sec:discussion}, we compared the observed emission-line ratios to different sets of photoionization and shock models from the literature. 
In particular, we considered:
\begin{itemize}
    \item Simple stellar population (SSP) models from \citet[][MP23 hereafter]{martinez-paredes2023}\footnote{Available in the Mexican Million Models Database: \url{https://sites.google.com/site/mexicanmillionmodels/}; \url{http://3mdb.astro.unam.mx:3686/}};
    \item SSP models with and without the ionizing output from X-ray binaries (XRBs) from \citet[][G24 hereafter]{garofali2024}\footnote{\url{https://github.com/kgarofali/sxp-cloudy}};
    \item Intermediate-massive black hole (IMBH) models from \citet[][R22 hereafter]{richardson2022}\footnote{\url{https://github.com/crichardson17/richardson_2022}};
    \item Shock models from \citet[][F24 hereafter]{flury2024}\footnote{\url{https://zenodo.org/records/14140949}} and the Mexican Million Models Shock Database\footnote{\url{http://3mdb.astro.unam.mx:3686/}}, described in \citet[][AM19 hereafter]{alarie2019}.
\end{itemize}
For all these models, we considered only the two lowest metallicity grids available (i.e., 5\% and 10\% Z$_\odot$ models for \citetalias{garofali2024}, \citetalias{richardson2022}, \citetalias{flury2024}, and 2\% and 10\% Z$_\odot$ for \citetalias{martinez-paredes2023}), in accordance with SBS\,0335-052~E's metallicity range \citep{papaderos2006,nakajima2024}.
Also, we took into account only bursts of star-formation with ages below 100~Myr (upper limit on age of SBS 0335-052~E) to match SBS 0335-052~E properties (\citealt{izotov1997,wofford2021,berg2022,mingozzi2024}, Martinez et al. in prep.). 
For all the grids, we consider the ionization parameter log($U$) between -4 and -1.5 (apart from \citetalias{richardson2022} where log($U$) reaches -0.5).
For \citetalias{martinez-paredes2023}, \citetalias{richardson2022} and shock models, we could also test the effect of different densities ($n_H$~$=10^2-10^4$~cm$^{-3}$ in \citetalias{martinez-paredes2023}, \citetalias{richardson2022}; $n_H$~$=1-10^4$~cm$^{-3}$ in shock models), finding small differences that only slightly impact the results shown in Sec.~\ref{sec:results}, as we discuss in Sec.~\ref{sec:discussion}. 
In the following, we give a summary of the characteristics of all these models. 

\citetalias{martinez-paredes2023} SSP models combine \textsc{CLOUDY} \citep{ferland2017} with the revised version of the \citet{bruzual2003} stellar population synthesis models introduced in \citet[][C\&B models]{plat2019}, spanning across several parameters (see their Tab.~4), including a \citet{kroupa2001} and top-heavy (x030) initial mass functions (IMFs) with a mass range up to 100 and 300 M$_\odot$. 
C\&B models accurately cover the evolution of O and B stars, including the Wolf-Rayet (WR) phase, whose contributions last a few Myr. C\&B models also include hot post-Asymptotic Giant Branch (pAGB) stars, also known as hot low-mass evolved stars (HOLMES; \citealt{stasinska2008}), which emit a strong UV continuum ionizing and exciting the surrounding medium at ages between 70 and 500 Myr (peak around 100 Myr). 

\citetalias{garofali2024} modeled simple stellar populations (SSPs) and a population of ultraluminous X-ray sources (ULXs), combining the ionizing impact of a SSP with XRBs, creating ``simple X-ray populations" (SXPs). 
As described in detail in \citetalias{garofali2024}, XRBs form in multiple generations following a burst of SF ($\gtrsim3-5$~Myr), thus can produce ionizing photons on longer timescales than single massive stars (up to $\sim20$~Myr). 
Also, when present, they dominate the X-ray power output of star-forming galaxies and have a luminosity that scales with SFR. 
\citetalias{garofali2024} used binary population synthesis models from \citet{fragos2013} and burst ages ranging from 1 to 20~Myr. 
For the SSPs, they used the stellar population synthesis code BPASS v2.2 \citep{eldridge2017}. 
Then, they created the SSP and SXP models with \textsc{CLOUDY} considering different values of gas-phase metallicities and ionization parameters, and a close geometry.
Concerning the chemical composition of the \textsc{CLOUDY} models, \citetalias{garofali2024} followed the prescriptions used in \citetalias{richardson2022}. 

\citetalias{richardson2022} made emission line predictions for $10^3-10^4-10^5-10^6$~M$_\odot$ IMBHs with \textsc{CLOUDY}, taking into account two extreme spectral energy distributions (SED) - a ``disk-plaw" and a ``qsosed" - for the incident radiation field (see \citetalias{richardson2022} Sec.~2.1) - and a starburst continuum including WR contribution, using the stellar population synthesis code BPASS~v2.0 \citep{stanway16} with a Kroupa IMF. 
To address different physical uncertainties, \citetalias{richardson2022} models at different values of metallicities and ionization parameters include both coincident and non-coincident mixing of the IMBH and starlight excitation, and open and closed geometries.
In particular, in this paper we show the 20~Myr instantaneous burst of star formation grids (in SSPs \oiii~$\lambda$5007/\hb\ peaks at 20~Myr; \citealt{xiao18}), with coincident mixing and different geometry.
The line ratios do not vary significantly in the non-coincident outputs, but can be slightly affected by the choice of geometry, as we comment in Sec.~\ref{sec:results}.

\citetalias{alarie2019} shock models are made with MAPPINGS-V \citep{sutherland2017}, a plane-parallel geometry and the same prescriptions of the well-known \citet{allen2008} models, but spanning a broad range of gas-phase metallicities and densities.
\citetalias{flury2024} models are made similarly with MAPPINGS-V, but uniformly implemented empirical abundance patterns and dust depletion treatment as well as updated atomic data.
Generally, two sets of models are taken into account: pure shocks and shocks+precursor. 
In this work, we consider only shocks+precursor models, since they also consider the gas entering the shock front.
The shown grids span values of shock velocity ($vs=100-1000$~km/s) and the magnetic field parameter ($B_0 = 10^{-4}-10$~$\mu$Gauss), which are the main parameters regulating the shock ionization spectrum and the effective ionization parameter. 
We find that \citetalias{alarie2019} and \citetalias{flury2024} shock models behave similarly in the range of parameters considered, thus in Sec.~\ref{sec:results} we show only \citetalias{flury2024} grids.

\section{Results}\label{sec:results}
\subsection{Emission-line maps}\label{sec:line-maps}
\begin{figure*}
    \includegraphics[width=1\textwidth]{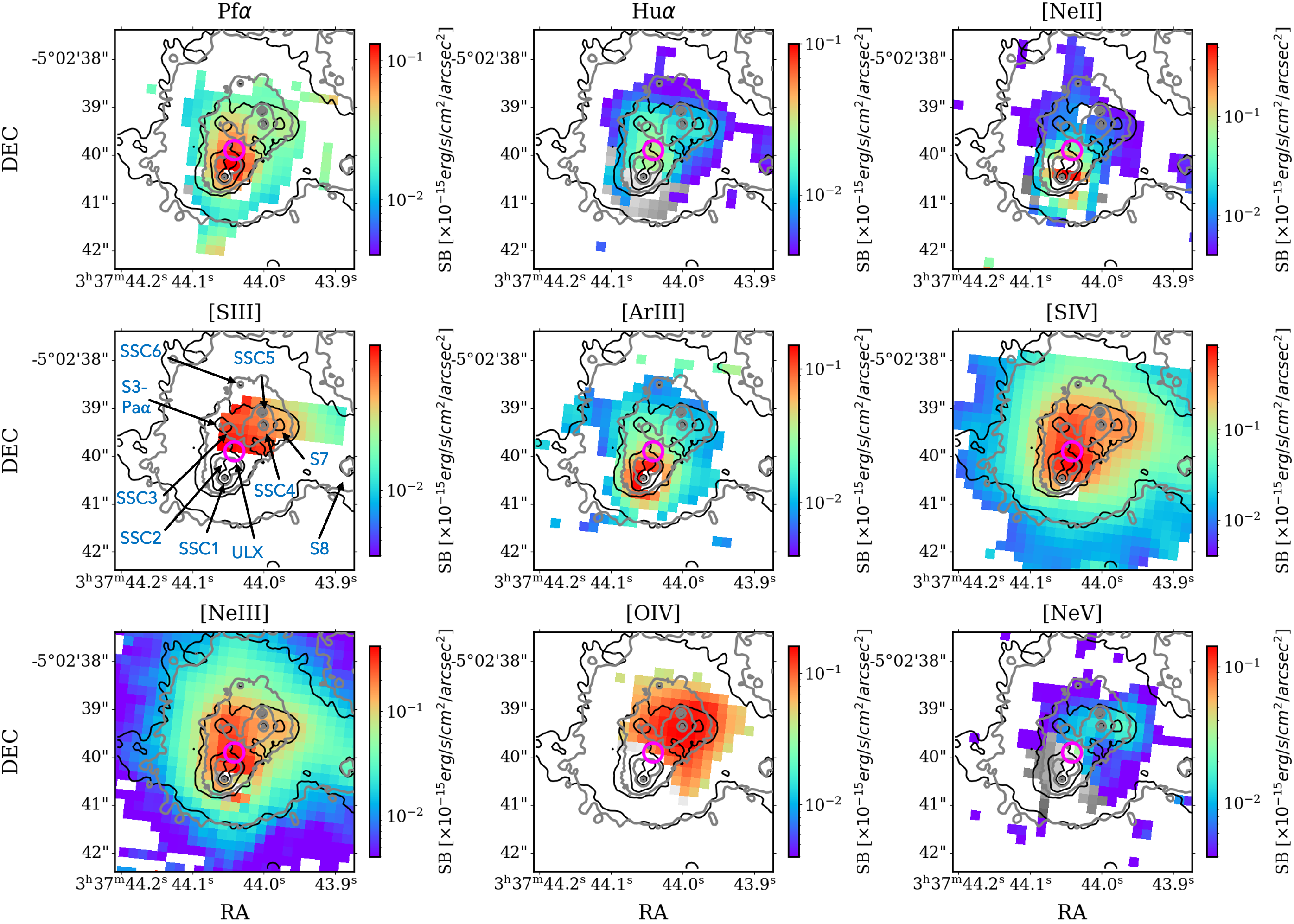}
\caption{Point-source subtracted emission-line maps corrected for dust attenuation on Channel 3 pixel scale (0.245"; 1"~$\sim280$~pc). All lines but \siii\ and \oiv\ are convolved to \neiii\ resolution ($FWHM_{PSF} \sim 0.6$"). The colored regions represent the measurements with $S/N>3$, while the gray areas have $S/N\sim2$. The black and gray contours show the H$\alpha$ and UV emission from the HST ACS WFC FR656N and F150LP filters (PID: 10575; 16209), respectively. The magenta circle with a diameter equal to Chandra’s positional uncertainty indicates the ULX position identified in this galaxy by Chandra \citep[][their Tab.~3]{prestwich2013}. The emission line maps are shown in order of increasing ionization potential (see Tab.~\ref{tab:listEmissionLines}). The positions of SBS~0335-052~E's several star-forming regions are shown on the \siii\ map (see also Fig.~\ref{fig:sbs-f140lp}). The extended high ionization region (bright \nev\ and \oiv) covers the majority of the stellar clusters and peaks around the S7 region. The FOV varies from 3.2"$\times$3.7", to 4.0"$\times$4.8", 5.2"$\times$6.2" and 6.6"$\times$7.7" for Channels 1 (\hpfa), 2 (\ariii, \siv), 3 (\hhuma, \neii, \nev, \neiii) and 4 (\siii, \oiv), respectively. }
\label{fig:emissionlines}
\end{figure*}

Fig.~\ref{fig:emissionlines} shows the emission line maps of the extended MIR emission lines obtained after the point source subtraction: \hpfa~$\lambda$~7.49$\mu$m, \hhuma~$\lambda$~12.37$\mu$m, \neii~$\lambda$~12.81$\mu$m, \siii~$\lambda$~18.71$\mu$m, \ariii~$\lambda$~8.99$\mu$m, \siv~$\lambda$~10.51$\mu$m, \neiii~$\lambda$~15.56$\mu$m, \oiv~$\lambda$~25.89$\mu$m, and \nev~$\lambda$~14.32$\mu$m, in order of ionization potential (see Tab.~\ref{tab:listEmissionLines}). 
The maps of all the lines but \siii\ and \oiv\ (Channel 4; $FWHM_{PSF}\sim 0.9$", $\sim 252$~pc) are convolved with the spatial resolution at the \neiii\ wavelength ($FWHM_{PSF}\sim 0.6$", $\sim 168$~pc) and rescaled to Channel 3 pixel scale (0.245").
The FOV varies from 3.2"$\times$3.7", to 4.0"$\times$4.8", 5.2"$\times$6.2" and 6.6"$\times$7.7" for Channel 1 (\hpfa), 2 (\ariii, \siv), 3 (\hhuma, \neii, \nev, \neiii) and 4 (\siii, \oiv), respectively. 
In the \siii\ map panel, we indicate the positions of several star-forming regions in SBS~0335-052~E as we did in Fig.~\ref{fig:sbs-f140lp}.

\hpfa\ and \hhuma\ maps follow the \ha\ distribution, which peaks in young star-forming regions ($t<3$~Myr), such as the position of SSC1 (point source) and SSC2 as well as S7 and S3-Pa$\alpha$, slightly shifted with respect to the other SSCs and very bright in Pa$\alpha$ \citep{thompson2009}.
\neiii\ and \siv\ have a very similar distribution with their emission extending across the MIRI/MRS FOV, due to their similar I.P. (see Tab.~\ref{tab:listEmissionLines}). 
Higher ionization emission such as \nev\ and \oiv\ mainly comes from the north-west portion of the galaxy, covering the older clusters SSCs3-4-5-6 ($t\sim7-15$~Myr) and peaks around the S7 position. 
The UV emission (HST ACS 150LP; Fig.~\ref{fig:sbs-f140lp}) shows a different pattern with respect to MIR and optical hydrogen lines, tracing the six SSCs and peaking in the \nev\ and \oiv\ high ionization region.
We highlight that \heii~$\lambda$4686 (similar I.P. to \oiv) emission studied by \citet{kehrig2018} with MUSE data closely follows the UV contours.

We notice that the peak of \neiii\ and \siv\ emission is close to the ULX position revealed by \citet[][their Table~3; see also \citealt{kehrig2018}]{prestwich2013} with Chandra (magenta circle). 
However, there is no clear correspondence between the ULX X-ray emission and high ionization lines, since the \nev, \oiv\ and UV emission peaks are shifted towards S7.
We highlight that the ULX extracted spectrum is also characterized by \feii~$\lambda$5.34$\mu$m emission, that is not shown in Fig.~\ref{fig:emissionlines} since it does not show clear extended emission (see Fig.~\ref{fig:psf-maps}). 

The region where we subtracted the point-source emission, located at the position of SSC1, shows either a reduced or tentative emission (gray regions with $S/N\sim 2$) in the line maps. 
This can be clearly seen in the \hhuma\ and \nev\ maps. 
This could be due to a real lack of extended emission around SSC1 or to uncertainties in the PSF subtraction procedure. 
As shown in Fig.~\ref{fig:1dspecpsf} the point source SSC1 shows almost all the emission lines reported in Fig.~\ref{fig:emissionlines}, apart from \nev\ and \arii. Finally, we extracted a spectrum centered at the point source position in the residual cube (i.e., after point-source subtraction), summing all the spaxels within one PSF FWHM, confirming that the absence of \nev\ is not due to residual fringing in Channel~3 (see Sec~\ref{sec:observations}).
The point source spectrum also shows clear \feii\ emission (see also Fig.~\ref{fig:psf-maps}). 

\subsection{Emission-line ratio maps}\label{sec:line-ratios}
\begin{figure*}
\centering
    \includegraphics[width=1\textwidth]{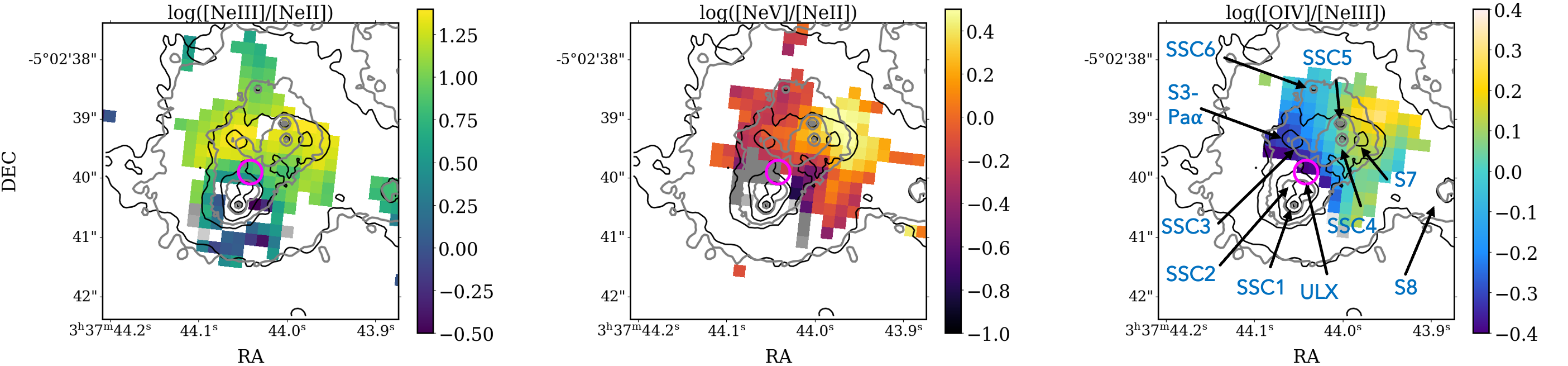}
\caption{\neiii/\neii, \nev/\neii\ and \oiv/\neiii\ line-ratio maps, where the colored spaxels have $S/N>3$ for the displayed emission lines. \neiii/\neii, tracing the ionization parameter, peaks around S3-Pa$\alpha$, while \nev/\neii\ and \oiv/\neiii, both tracing the radiation hardness, peak close to S7. The gray areas show regions where \neii\ and \nev\ have $S/N\sim2$.}
\label{fig:lineratios}
\end{figure*}
Fig.~\ref{fig:lineratios} shows \neiii/\neii, \nev/\neii\ and \oiv/\neiii\ line ratio extended emission maps with a color-coding chosen to enhance the peak regions to aid with the interpretation of the diagnostic diagrams.
\neiii/\neii\ is known to be a good tracer of the ionization parameter log($U$), while \nev/\neii\ and \oiv/\neiii\ are found to be sensitive also to the hardness of the ionizing spectrum \citep{richardson2022,garofali2024}.
Indeed, these line ratios do not show the same behavior.
The peak of \nev/\neii\ and \oiv/\neiii\ maps are located around region S7, aligning with the peak in \ha\ emission (see black contours). 
While \neiii/\neii\ is also enhanced in S7, it instead peaks at the position of S3-Pa$\alpha$, also bright in \ha\ (see black contours). 
SSC3, SSC4, SSC5 and SSC6 are characterized by high ionization emission, but lower \nev/\neii\ and \oiv/\neiii\ line ratios than S7. 
As mentioned above, the extended region covering SSC1 (point source) and SSC2 does not have sufficient S/N in \neii, \nev\ and \oiv. 
Finally, S8 does not have either \nev\ or \oiv\ emission but shows intermediate values of \neiii/\neii. Tab.~\ref{tab:compilationResults} reports the line ratios for the different regions discussed above.
\subsection{Ionized gas kinematics maps}\label{sec:kinmaps}
\begin{figure*}[htp]
\centering
    \includegraphics[width=0.33\textwidth]{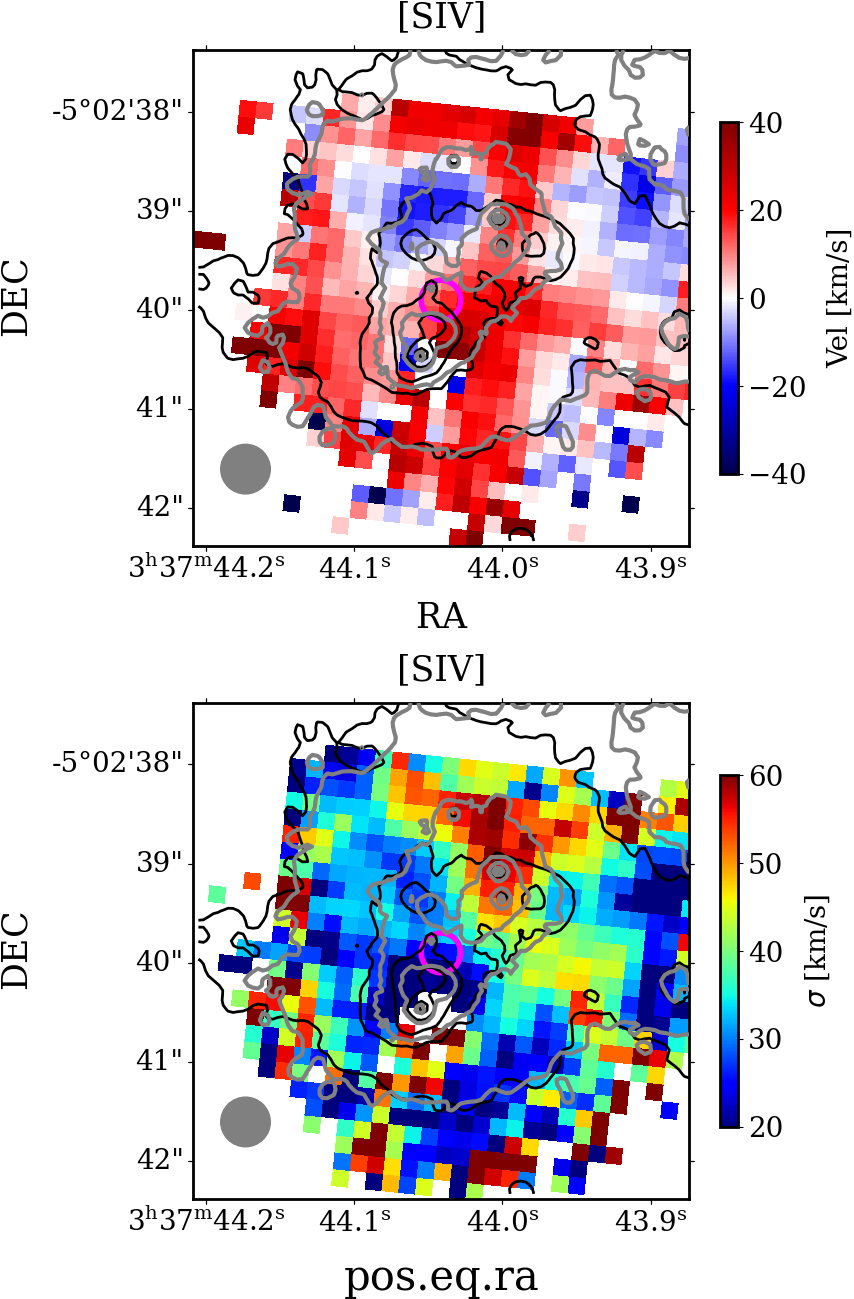}
    \includegraphics[width=0.33\textwidth]{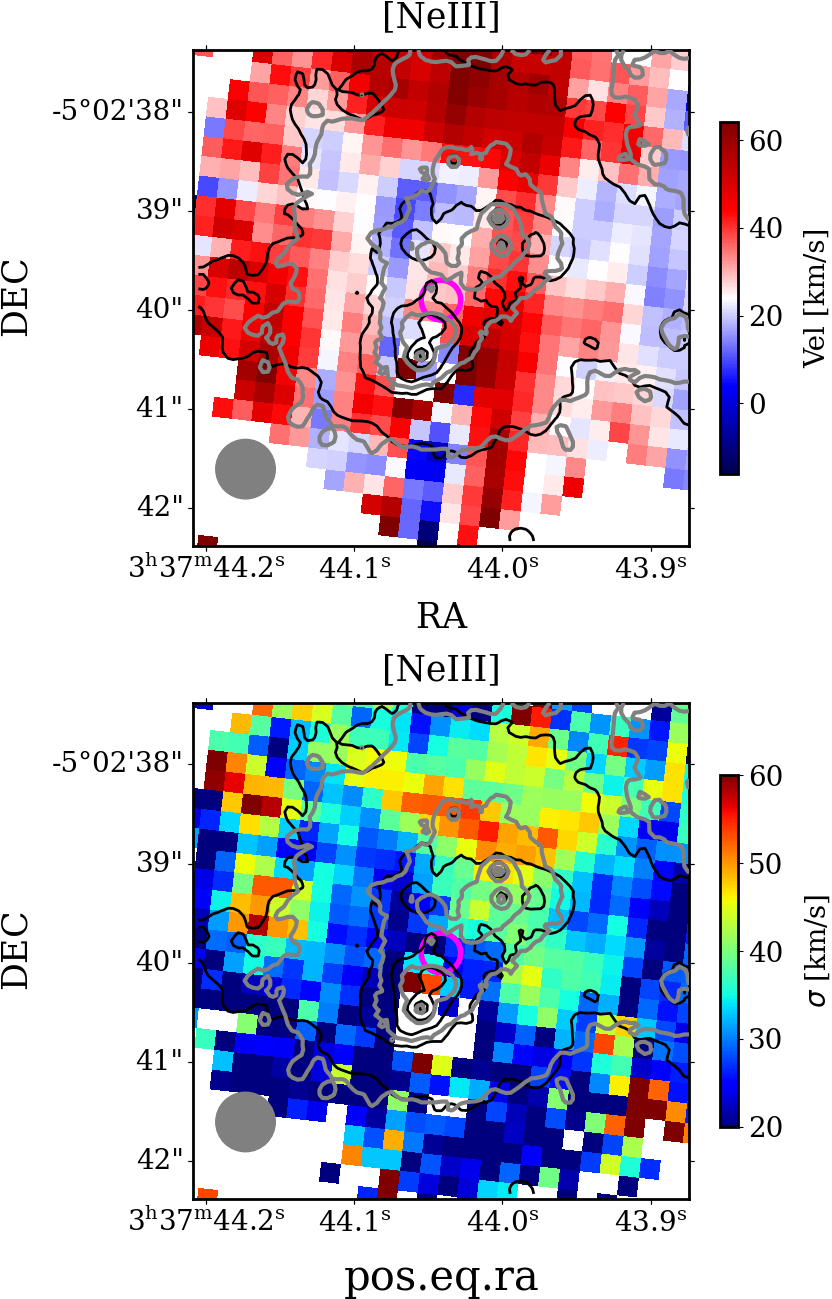}
    \includegraphics[width=0.33\textwidth]{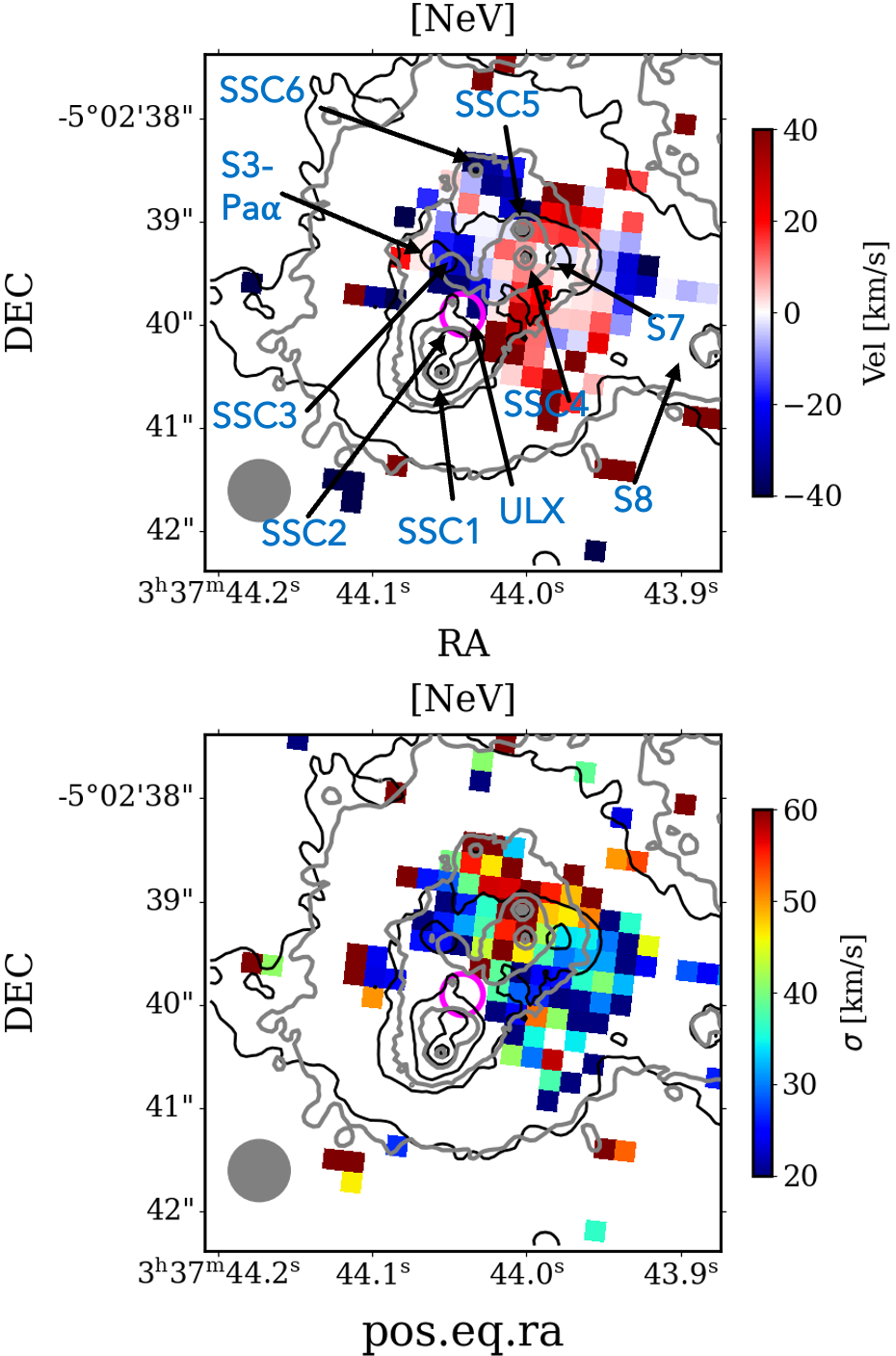}
\caption{Velocity (top panels) and velocity dispersion (corrected for instrumental broadening; bottom panels) maps of the \siv, \neiii\ and \nev\ emission lines, in order of ionization potential (see Tab.~\ref{tab:listEmissionLines}), with the same orientation, labeling and contours as Fig.~\ref{fig:emissionlines} (1"~$\sim280$~pc). The velocity is centered on the systemic velocity of the galaxy (see footnote~\ref{myfootnote}). We show only the spaxels with $S/N>3$. \siv\ belongs to Channel 2, \neiii\ and \nev\ to Channel 3, and the maps are not convolved and are in their native pixel scale. The velocity maps show a similar pattern with blueshifted emission around the S3-Pa$\alpha$ source and on the west part of the FOV, and redshifted emission from north to south and in the east part of the FOV, possibly as a result of an expanding shell (see \citealt{izotov2006}). The velocity dispersion maps show an enhancement in the north region (``V" shape) with values North from SSC4-5-S7, where we see the high ionization \oiv\ and \nev\ emission, with a slight shift to the north in the lower ionization lines. The gray circle shows the $FWHM_{PSF}$ ($\sim 0.5$" for \siv; $\sim 0,6$" for \neiii\ and \nev) to highlight that we resolve the displayed velocity pattern and high $\sigma$ features.}
\label{fig:kinemissionlines}
\end{figure*}
Fig.~\ref{fig:kinemissionlines} shows the \siv, \neiii\ and \nev\ velocity and intrinsic velocity dispersion maps. 
The velocity maps are centred on the systemic velocity of this galaxy according to previous optical studies ($v_0 = cz\sim4503$~km/s; \citealt{moiseev2010})\footnote{\label{myfootnote} We centred the \neiii\ velocity map -24~km/s below the systemic velocity $v_0= cz = 4053$~km/s found by \citet{moiseev2010}. In particular, we find this specific systematic shift in \hhuma\ and \neii\ (Ch.~3A), and \neiii\ (Ch.~3C) velocity maps also obtained for the Stage 3 products, so it does not depend on the point-source subtraction procedure explained in Sec.~\ref{app:psf-sub}. We suspect that this shift is not connected to the physical conditions of the gas, as it corresponds to $\sim$~half the wavelength step in Channel~3. Also, \hpfa, \ariii\ and \oiv, located in different channels and characterized by different I.P., show consistent velocity maps to \siv\ and \nev.}, while the intrinsic velocity dispersion maps are obtained by subtracting in quadrature the instrumental broadening dependent on wavelength from the observed velocity dispersion (see Eq.~1, \citealt{jones-miri2023}).

The MIRI/MRS FOV is limited to the central $\sim1$~kpc$^2$ of SBS~0335-052~E and does not cover the entire ionized gas distribution traced in \ha\ by previous studies (e.g., \citealt{moiseev2010,herenz2023}), with signs of outflowing gas extending up to $\sim15$~kpc \citep[][in the north-west direction, along the galaxy minor axis]{herenz2023}. 
Thus, despite \citet{moiseev2010} showing that on larger scales there is a disk-like north-west to south-east gradient, we only see the inner regions known to be characterized by perturbed kinematics \citep{izotov2006,moiseev2010,herenz2017,herenz2023}. 
In particular, \siv, \neiii\ and \nev\ maps trace progressively higher ionization gas (Tab.~\ref{tab:listEmissionLines}) and show a consistent velocity pattern, with a blueshifted emission around the position of S3-Pa$\alpha$ and in the north-west portion of the MIRI/MRS FOV (on the right of S7 and at the top of S8), and redshifted emission from north to south and in the east region. 

Despite the kinematic complexity shown in the velocity maps, the emission lines are well-fitted with a single Gaussian, broadening in the north-west region of the MIRI/MRS FOV.
Interestingly, the \siv, \neiii\ and \nev\ intrinsic velocity dispersion maps seem to peak in slightly shifted positions. 
They all show a curious ``V" shape (less visible in \nev) with enhanced $\sigma$ that peaks in SSCs 4-5 in the \nev\ map, $\sim1-2$~px north than that in \neiii\ and further north in \siv. 
We notice that the region with enhanced $\sigma$ in the \nev\ map is located close to the peak of the \nev/\neii\ and \oiv/\neii\ line ratios (Fig.~\ref{fig:lineratios}) tracing the hardening of the radiation field.

The regions where \neiii\ and \siv\ $\sigma$ peaks probably have no \nev\ emission due to their lower ionization (they also partly show \neii\ emission; Fig.~\ref{fig:emissionlines}) and are characterized by enhanced redshifted velocities, possibly showing a flow of gas moving away from the \nev\ high ionization region.
This picture is consistent with what found by \citet{izotov2006} with high-resolution ($R\sim10,000$) VLT/GIRAFFE ARGUS spectra, showing double-peaked \ha\ profiles in the region north of SSCs 4-5 (see their Fig.~7a and 8).
In particular, they interpreted these features with the presence of an expanding shell of ionized gas at $\sim 50$~km/s, while the \ha\ broadening in SSCs 4-5 with higher dynamic activity than the surroundings. 

From a visual comparison of the maps, the 15~kpc outflow revealed by \citet{herenz2023} in \ha\ is not clearly connected to the central kpc kinematics and the location of the sources S7, SSC4-5. 
Interestingly, previous optical integral-field spectroscopy studies have also revealed an offset between the expanding superbubble shells and the star clusters driving the global outflow in galaxies similar to SBS~0335-052~E (e.g., \citealt{martin2024}).
Finally, we stress that no ionized outflow is detected in SBS~0335-052~E HST COS UV absorption lines in correspondence of the sources S7, SSC4-5 \citep{xu2022,parker2024}. 

\subsection{Diagnostic diagrams}\label{sec:diag-ratios}
\begin{figure*}[tbp]
\centering
    \includegraphics[width=.99\textwidth]{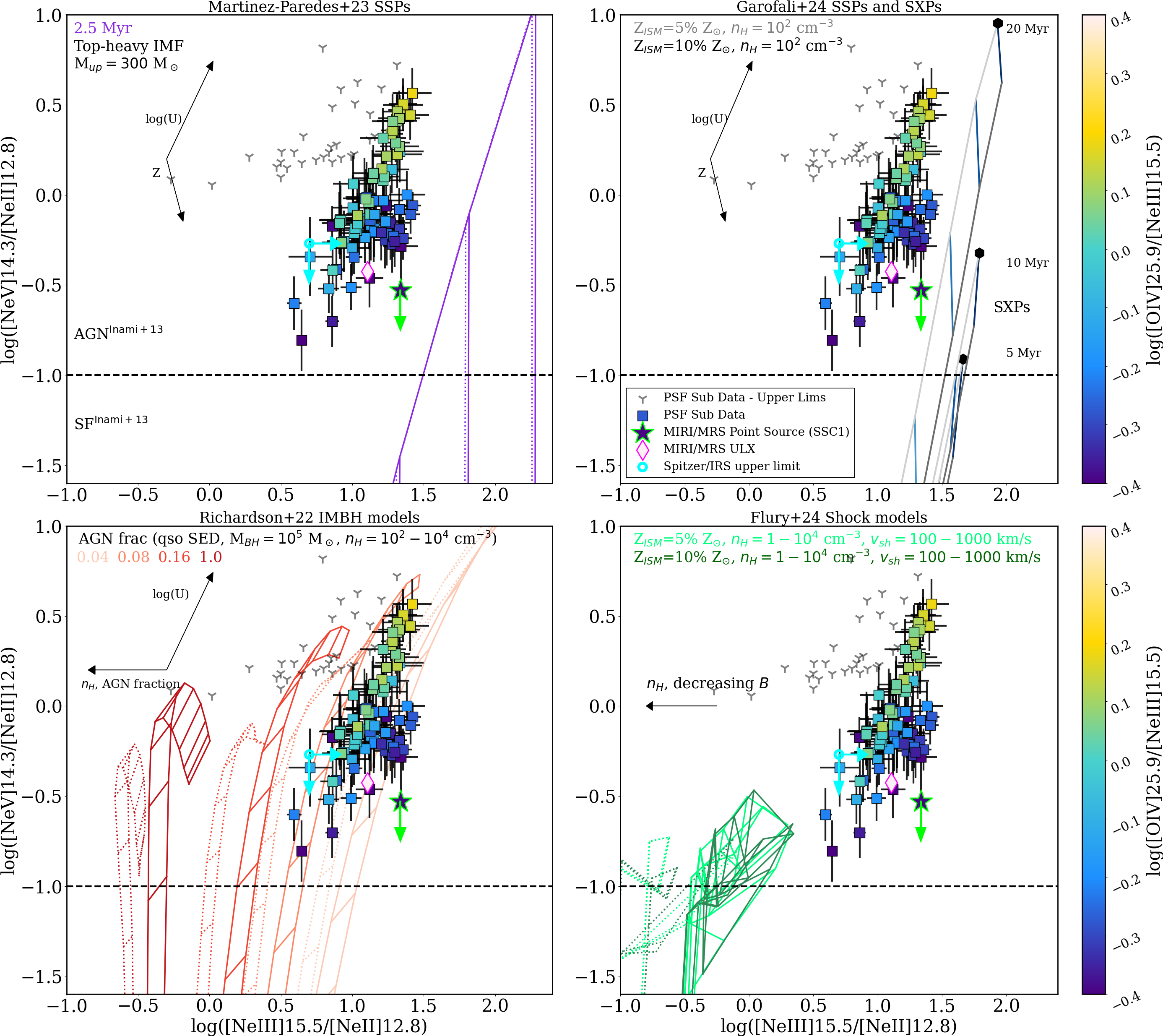}
\caption{\nev/\neii\ versus \neiii/\neii\ diagnostic diagram color-coded as a function of \oiv/\neiii\ line ratio, with overplotted the four sets of low-metallicity models presented in Sec.~\ref{sec:models}: 2.5 Myr SSPs models from \citetalias{martinez-paredes2023}, 5, 10 and 20 Myr SSPs and SXPs models from \citetalias{garofali2024}, IMBH models ($M_{BH}=10^5$~M$_\odot$; 20 Myr SSP; 4\%, 8\%, 16\%, 100\% AGN fraction; closed geometry) from \citetalias{richardson2022} and shock+precursor models from \citetalias{flury2024}. The model details are indicated in the different panels. The dotted grids have higher density. The star and diamond symbols in the scatter plots show the line ratios measured for the point source (SSC1) and the ULX spectrum, respectively. The cyan point indicates the Spitzer upper limit (\nev\ and \neii\ fluxes are upper limits; \citealt{houck2004,hao2009}). The black dashed line shows the criterion to separate SF from AGN from \citet{inami2013}. The gray points are upper limits ($S/N\sim$; see gray regions in Fig.~\ref{fig:lineratios}), tracing extended emission close to the point source. The color-coding is the same as Fig.~\ref{fig:lineratios}, to visually understand to which regions the line ratios correspond. The models that show better agreement with the measurements are \citet{richardson2022} IMBH models with a low AGN fraction ($4-8$\%). 
This diagnostic diagram is only marginally affected by the wavelength-dependent MIRI PSF, pixel scale, and dust reddening since all the lines belong to Channel~3, where the attenuation law is flat. }
\label{fig:diags1}
\end{figure*}
In this paper, we explore the following diagnostic diagrams widely used in the literature (see \citealt{hao2009,weaver2010,inami2013}, but also \citetalias{richardson2022}, \citetalias{martinez-paredes2023}, \citetalias{garofali2024}): 
\begin{itemize}
    \item \nev/\neii\ versus \neiii/\neii, color-coded as a function of \oiv/\neiii\ (Fig.~\ref{fig:diags1});
    \item \neiii/\neii\ vs \siv/\neii, color-coded as a function of \oiv/\neiii\ (Fig.~\ref{fig:diags2}, App.~\ref{app:other-diags});
    \item \neiii/\neii\ versus \oiv/\neiii, color-coded as a function of \nev/\neii\ (Fig.~\ref{fig:diags3}, App.~\ref{app:other-diags}). 
\end{itemize}
We highlight that \nev/\neii\ versus \neiii/\neii\ represents the most reliable diagnostic diagram explored here, since these lines lie in Channel 3, where the attenuation law is flat, thus they are minimally impacted by PSF variations and dust correction. 
\siv/\neii\ instead can be affected by dust reddening, while \oiv/\neiii\ is the ratio most affected by the wavelength-dependent MIRI/MRS PSF variation correction, as well as residual fringing, given that \neiii\ is located in Channel~3c and \oiv\ in Channel~4c. This is also likely affecting the \oiv/\neiii\ value estimated for the ULX extracted spectrum, that is $\sim0.5$~dex higher than the median value of the spatially resolved spaxels (see Fig.~\ref{fig:diags3}, Tab.~\ref{tab:compilationResults}).
For these reasons, we mainly focus this section on \nev/\neii\ versus \neiii/\neii, moving the other two diagnostic diagrams in App.~\ref{app:other-diags}.

The color-coding of each diagram corresponds to its respective map shown in Fig.~\ref{fig:lineratios}, to visually understand to which regions the shown line ratios correspond.
We decided to focus on these diagnostic diagrams for the following reasons: 
(i) \neiii/\neii\ line ratio traces log($U$) and it can also be a good diagnostic of AGN fraction for IMBHs (\citetalias{richardson2022}). 
In general, it can widely vary in galaxies, offering a good opportunity to investigate their properties.
(ii) \siv/\neii\ is known to behave similarly to \neiii/\neii, given the common denominator and \siv\ and \neiii\ similar I.P. 
Indeed, \citet{groves2008} found a very tight linear correlation between these line ratios that holds - with some scatter - for very different galaxies, including starbursts, ultra-luminous infrared galaxies, AGN, and also BCDs such as SBS~0335-052~E. 
As such, we can use these line ratios to test how SBS 0335-052~E compares to similar BCDs.
(iii) \oiv\ and \nev\ instead are high ionization emission lines (see Tab.~\ref{tab:listEmissionLines}), usually considered good indicators of ionizing mechanisms beyond SF, such as AGN and IMBHs. 
In particular, as mentioned in Sec~\ref{sec:line-ratios}, both \oiv/\neiii\ and \nev/\neii\ are able to trace the hardness of the ionizing spectrum (\citetalias{richardson2022,garofali2024}).
For instance, a typical criterion to differentiate AGN from SF dominated galaxies is log(\nev/\neii)~$>-1$, given the much lower ionization potential of \neii, abundantly produced in \hii\ regions (e.g., \citealt{inami2013}). 
Also, \nev\ is usually undetected in pure-star-forming regions, and difficult to reproduce even with SXP models (e.g., only at specific log($U$) and $Z$; see \citetalias{garofali2024}). 
A criterion to distinguish SF and IMBHs ionization was proposed by \citet{richardson2022}, introducing separators according to \oiv/\neiii\ and \neiii/\neii\ line ratios.

Fig.~\ref{fig:diags1} (same for Fig.~\ref{fig:diags2} and Fig.~\ref{fig:diags3}, App.~\ref{app:other-diags}) show four panels where we overlay the diagnostic diagrams with a different set of low-metallicity models (see Sec.~\ref{sec:models}): 
\begin{itemize}
    \item \citetalias{martinez-paredes2023} 2.5~Myr SSPs, with top-heavy IMFs and $M_{up}=300$~M$_\odot$ (upper left panel); 
    \item \citetalias{garofali2024} SSPs and SXPs grids at 5, 10 and 20~Myr (upper right; the hexagon represents the highest line ratio each grid can reproduce);
    \item  \citetalias{richardson2022} IMBHs grids with $M_{BH}=10^5$~M$_\odot$, qso SED and 20 Myr SSP, with closed geometry and 4, 8, 16 and 100\% AGN fraction, highlighting that beyond 16\% the SED is completely dominated by the IMBH (lower left panel); 
    \item \citetalias{flury2024} shock models with precursor (lower right panel). 
\end{itemize}

It should be noted that for the \citetalias{martinez-paredes2023} SSP grids, the low metallicity 2.5 and 100~Myr SSPs show the highest line ratios for young and old stars, respectively. These two grids are the only ones (partially) able to cover the x- and y- ranges shown. 
However, we discarded the 100~Myr grid because old stars struggle to explain the observed number of ionizing photons observed in this system, as we discuss in Sec.~\ref{sec:disc-stars}. 
Also, we notice that the choice of the IMF does not affect the line ratios, but the choice of $M_{up}=300$~M$_\odot$ strongly enhances \nev/\neii.
All the other \citetalias{martinez-paredes2023} grids and \citetalias{garofali2024} SSPs and SXPs grids below 5~Myr fall outside the shown x- and y-ranges. 
\citetalias{richardson2022} IMBHs grids with $M_{BH}=10^{3-4}$~M$_\odot$ are also not shown because they cannot reproduce log(\neiii/\neii) higher than $\sim0-0.5$. The grids with $M_{BH}=10^{6}$~M$_\odot$ instead can overlap to the observed line ratios at 8\% AGN fraction, but at lower/higher fractions, they predict larger/lower \neiii/\neii. 

Fig.~\ref{fig:diags1} (same for \ref{fig:diags2} and \ref{fig:diags3}) scatter plots show the extended emission measured with our data (down-pointing arrows for upper limits), as indicated in the legend.
We stress again that the point source (star symbol) has only tentative \nev\ with $S/N\sim2$.
The cyan point indicates the \nev\ and \neii\ upper limits and the \neiii\ and \oiv\ detections from Spitzer flux measurements \citep{houck2004,hao2009}, which cover the entire region covered by MIRI/MRS. In particular, \citet{houck2004} reported that the Spitzer image had a prominent diffraction ring and was thus indistinguishable from the image of a point source (possibly SSC1 as we reveal in our data). 

From a quick look at Fig.~\ref{fig:diags1} (same for Fig.~\ref{fig:diags2}, App.~\ref{app:other-diags}), all SBS 0335-052~E MIR line ratios lie outside of the ``canonical" SF locus, ending in the ``AGN" designated region according to \citet{inami2013} and \citet{richardson2022} criteria (see also e.g., \citealt{goold2024}). 
This is already suggested by Spitzer upper limits (cyan point), but with MIRI/MRS data we can tell much more about the conditions of the ionized gas on a spatially resolved basis. 
Clearly, the \neiii/\neii\ and \nev/\neii\ diagnostic diagram disentangles increasing ionization parameter and radiation hardness sequences, with both \neiii/\neii\ and \nev/\neii\ progressively increasing with the log($U$) of different models (i.e., following the log($U$) arrow shown in the various panels).
\nev/\neii\ and \oiv/\neiii\ (color-coding), being more sensitive to the radiation hardness, increase also along the perpendicular direction of the increasing log($U$) arrow, at decreasing \neiii/\neii\ values. 
This effect is reproduced by harder ionization in the models, such as lower gas metallicity or older star ionization in SSPs models, and higher AGN fractions in IMBH models.

Another important takeaway from the top two panels of Fig.~\ref{fig:diags1} (same for Fig.~\ref{fig:diags2} and \ref{fig:diags3}, App.~\ref{app:other-diags}) is that all the low-metallicity young SSPs grids struggle to reproduce the observed MIR emission line ratios of SBS~0335-052~E.
This suggests the need for a further ionization source. However, SXP models, including the contribution of X-ray binaries, show similar limitations. 
The only models that can reproduce SBS~0335-052~E MIR \nev/\neii, \neiii/\neii\ and \siv/\neii\ line ratios are models including the contribution of an IMBH ($4-8$\%), spanning a range of log($U$).
However, we highlight that these IMBH models still struggle to reproduce the highest \oiv/\neiii\ line ratios (Fig.~\ref{fig:diags3}). 
\siv/\neii\ vs \neiii/\neii\ ratios observed with Spitzer in BCDs are consistent with the empirical relation from \citealt{groves2008}, shown as the magenta line in Fig.~\ref{fig:diags2}. Since low-metallicity SSP, SXP, and shock models fail to reproduce this relation, they also struggle to match the lower-ionization MIR lines (below \nev) typically observed in BCDs.

Given the critical densities of the MIR lines, another property that can affect the parameter space covered by these line ratios is the gas density. 
For instance, \nev~$\lambda$14.3 is part of a doublet, showing a second transition at 24.3~$\mu$m, with lower critical density ($n_{crit}=3\times10^4, 5\times10^5$~cm$^{-3}$ at 24.3, 14.3~$\mu$m). This makes increasing \nev~$\lambda$14.3/\nev~$\lambda$24.3 ratios a good electron density ($n_e$) diagnostic in the range $n_e = 10^2 - 5\times10^5$~cm$^{-3}$ (e.g., \citealt{fernandez-ontiveros2016}). 
We do not detect \nev~$\lambda$24.3, which could be either due to the worsening of fringing in Channel~4 or to a high density of the \nev\ emitting region. 
We favor the latter explanation given the fact that the \nev\ emitting region shows high density according to previous UV and optical studies (i.e., $n_e$(\feiii)~$= 2440 \pm 506 $; $n_{e}$(\ciii)~$= 4258\pm1067$; $n_e$(\ariv)~$= 1107 \pm 5 $; $n_e$(\niv)~$< 42000 $; \citealt{mingozzi2022}). 
Interestingly, only models with $n_e\sim10^{3.5}-10^4$~cm$^{-3}$ seem able to reproduce a \nev~$\lambda$14.3 stronger than \nev~$\lambda$24.3. 
However, these high-density grids are slightly shifted towards lower \neiii/\neii\ and \oiv/\neiii\ line ratios (dotted grids in all diagrams), and therefore tend to make the models less consistent with the data.

Overall, all these considerations suggest that none of the current state-of-the-art models can perfectly match our observed MIR line ratios.
In Sec.~\ref{sec:discussion}, we discuss the feasibility of these results, also comparing them with previous spatially resolved studies performed on this galaxy. 


\section{Discussion}\label{sec:discussion}
As shown in the previous sections (see also Fig.~\ref{fig:sbs-f140lp}, right panel), the MIRI/MRS data of SBS 0335-052~E are characterized by a bright point source emission, clearly showing the instrument PSF structure. 
As explained in Sec.~\ref{sec:data-analysis} and App.~\ref{app:psf-sub}, we were able to model the point source contribution and accurately separate it from the underlying extended emission.
Tab.~\ref{tab:compilationResults} shows a summary of the information that MIRI/MRS can give us on the several star-forming regions of SBS~0335-052~E, as well as information gathered from the many previous works in the literature on this intriguing object to help with the interpretation of our data and the models presented in Section~\ref{sec:models}.
In the following subsections, we discuss the origin of the point source and extended emission, their main ionization sources and we connect our results with the high-$z$ universe.
\begin{table*}
    \begin{center}
    \begin{tabular}{lccccc} 
            Region & Age [Myr]    & Mass [$M_\odot$]                    & 12+log(O/H)                & \multicolumn{2}{c}{Notable features}  \\
        \hline 
        SSC1   & 3.0 &  4.7$\times 10^5$ & \multirow{2}{*}{7.31}         &                   \multicolumn{2}{l}{Point source in MIRI-MRS and NICMOS 1.6~$\mu$m (T09)} \\ 
        SSC2   & 3.0 &  3.7$\times 10^5$ &                               &     \multicolumn{2}{l}{SSC1-2 uniquely has dust and H$_2$, and thermal radio emission (J09)} \\ 
        SSC3   & 7.0   &  7.1$\times 10^5$ &       7.31             &  \multicolumn{2}{l}{WR Blue bump (P06; K18)} \\ 
        SSC4   & 11.0  &  1.1$\times 10^6$ & \multirow{2}{*}{7.27}  &  \multicolumn{2}{l}{SSCs4-5 characterized by $\sigma$(\nev) peak and tentative radio emission (J09)} \\ 
        SSC5   & 13.0  &  2.9$\times 10^6$ &                               & \multicolumn{2}{l}{Point source in NICMOS 1.6~$\mu$m (T09)} \\ 
        SSC6   & 11.0  &  2.6$\times 10^5$ &           ---                    & \multicolumn{2}{l}{Point source in NICMOS 1.6~$\mu$m (T09); $\sigma$(\siv) and  $\sigma$(\neiii) peak} \\ 
        S3-Pa$\alpha$&  $10$~(T09)    &          2.3$\times10^6$~(T09)         &          7.38                     & \multicolumn{2}{l}{Bright H lines (T09)} \\ 
        S7     &    $4$~(T09)        &         9.4$\times10^4$~(T09)         &         7.10                      & \multicolumn{2}{l}{Bright H lines (T09); \nev/\neii\ and \oiv/\neiii\ peak} \\ 
        S8     &     $8$~(T09)       &       2.2$\times10^5$~(T09)         &        7.21            & \multicolumn{2}{l}{$\times$}                            \\ 
        ULX    &      ---      &       ---            &         ---                      & \multicolumn{2}{l}{Revealed by P13 (see also K18)} \\ 
        \hline
        Region &\neiii/\neii & \nev/\neii & \oiv/\neiii & \siv/\neii & \nev~$\lambda$14.3$\mu$m           \\
        \hline
        SSC1$^\dagger$     & $1.34\pm0.03$  & $<-0.53$  & $-0.40\pm0.07$  & $1.71\pm0.03$ &  $\times$  \\ 
        SSC2               & $-0.13\pm0.01$ & $<-1.49$  & $-0.7\pm0.5$ & $0.546\pm0.004$ & $\times$  \\ 
        SSC3+S3-Pa$\alpha$ & $1.36\pm0.01$  & $-0.14\pm0.02$ & $-0.29\pm0.01$ & $1.64\pm0.01$ &  \checkmark \\ 
        SSC4-5             & $1.19\pm0.04$  & $0.16\pm0.04$  & $0.05\pm0.01$ & $1.64\pm0.03$ & \checkmark \\ 
        SSC6               & $0.85\pm0.04$  & $-0.20\pm0.04$ & $-0.07\pm0.02$ & $1.18\pm0.03$ & \checkmark \\ 
        S7                 & $1.31\pm0.02$  & $0.35\pm0.02$  & $0.11\pm0.01$ & $1.66\pm0.02$ & \checkmark \\ 
        S8                 & $0.60\pm0.08$  & ---   & --- & $1.08\pm0.04$ & $\times$     \\ 
        ULX$^\dagger$      & $1.11\pm0.02$ & $-0.43\pm0.04$ & $0.48\pm0.02$ & $1.14\pm0.02$ &\checkmark \\ 
        \hline
       
    \end{tabular}
    \end{center}
    \caption{Ages and masses from \citet{reines2008,adamo2010}, unless specified otherwise; 12+log(O/H) from \citet[][]{thuan2004,papaderos2006} with uncertainties of $\sim0.02$~dex; other properties from \citet[][P06,T09,J09,P13,K18]{papaderos2006,thompson2009,johnson2009,prestwich2013,kehrig2018} and this work. All MIR line ratios are in logarithmic format. Line fluxes have been measured on the maps using apertures whose FWHM matches the PSF value, except for SSC1 and ULX (marked with $\dagger$) for which the fluxes have been measured on their integrated spectra.}
    \label{tab:compilationResults}
\end{table*}
\subsection{The bright point source, SSC1}\label{sec:point-source}
As introduced in Sec.~\ref{sec:data-analysis}, the point source is the source SSC1 (Fig.~\ref{fig:sbs-f140lp}).
SSC1 appears as a point source in the MIRI/MRS data, possibly because of its dusty nature, which makes it particularly bright in the MIR. 
As described in Sec.~\ref{sec:dust}, when analyzing the MIR continuum, we find a very high attenuation for the point source spectrum ($A_V\sim15$). 
Indeed, as shown in Fig.~\ref{fig:1dspecpsf}, this spectrum is characterized by the silicate feature at 9.7~$\mu$m, which is much fainter in the other regions of the galaxy. 
This is consistent with the fact that \citet{hunt2014} revealed dust continuum emission uniquely coming from the region around SSC1 using ALMA data (see their Fig.~1).
This region is also uniquely characterized by $H_2$ emission, tracing warm molecular gas, as first revealed by NICMOS data \citep{reines2008,thompson2009}. 
We confirm this with our data, observing several H$_2$ rotational transitions around SSC1 (from S(1) to S(7)), which will be modeled and discussed in detail in a forthcoming paper (del Valle-Espinosa et al., in prep.). 
However, the $H_2$ emission may come from a more extended region, given that the H$_2$ spatial distributions do not show the PSF pattern \citep[see also][their Fig.~3]{thompson2009}.
The point source region and close surroundings also show high \ha\ and Pa$\alpha$ equivalent widths \citep{reines2008} and purely thermal radio free-free emission, which has been interpreted as due to active star-formation in a dense environment (i.e., $\sim12,000$~O7.5~V stars, $n_e\sim 10^3-10^4$~cm$^{-3}$, $SFR\sim$1.3~M$_\odot$/yr$^{-1}$; \citealt{johnson2009}). 

It should be noted that SSC1 also displayed typical PSF Airy ring in HST Near Infrared Camera and Multi-Object Spectrometer (NICMOS) NIR data, tracing Pa$\alpha$ and 1.6~$\mu$m continuum emission (0.2"/px, comparable to MIRI/MRS) in \citet{reines2008,thompson2009}. 
In particular, \citet{thompson2009} showed that high-resolution (0.075~"/px) NICMOS camera 2 data could resolve the SSC1 emission and estimated the point source emitting region to be $\sim18$~pc.
\citet{thompson2009} showed that also SSC5 and SSC6 show a point source structure in the NIR NICMOS 1.6~$\mu$m continuum, with SSC5 seeming so compact (spatial extension of $\lesssim 10$~pc) that could not be resolved even in their high-resolution NICMOS dataset. 
In our MIRI/MRS data, we do not reveal any point source emission corresponding to SSC5 and SSC6, implying that those regions of the galaxy are instead dominated by extended emission in the MIR. 

All the elements listed above seem consistent with SSC1 being a very young embedded star-forming region \citep{hunt2001,reines2008,adamo2010}.
Indeed, as shown in Fig.~\ref{fig:diags1} and ~\ref{fig:diags3}, a subset of 2.5~Myr SSP models from \citetalias{martinez-paredes2023} could tentatively explain the point source \neiii/\neii\ and \oiv/\neiii\ line ratios - despite the fact that they fall in the ``AGN" regions of the diagnostic diagrams - and possibly the \nev/\neii\ upper limit. 
However, SSP models, are not capable of reproducing the corresponding lower ionization \siv/\neii\ line ratio (see Fig.~\ref{fig:diags2}), which seems more consistent with \citetalias{richardson2022} $M_{BH}=10^5$~M$_\odot$ grids with an AGN fraction below 8\%. 

Interestingly, the point source spectrum is also characterized by \feii\ emission, mainly located around SSCs1-2. \feii\ is considered an important shock tracer, because of (i) its many levels with low excitation energies easily excited in shocked gas, (ii) its low ionization potential (Tab.~\ref{tab:listEmissionLines}) that can make it ionized by FUV radiation from the shock front in the neutral H gas and (iii) is enhancement could be due to grain destruction by shocks (e.g., \citealt{koo2016} and references therein). 
However, 5-10\%~Z$_\odot$ shock models cannot reproduce the observed MIR line ratios.

Finally, as we will discuss in detail in Sec.~\ref{sec:extended-emission}, \citet{kehrig2018} derived SBS~0335-052~E \heii~$\lambda$4686-ionizing budget (including at the position SSCs1-2; knot~B in their nomenclature), finding that it can only be produced by either single, rotating metal-free stars or a 0.05\%~Z$_\odot$ binary population and a top-heavy IMF.
Since \heii\ and \oiv\ have similar I.P., they are likely powered by the same ionizing source, which could also cause the lower ionization MIR emission.
However, a 0.05\%~Z$_\odot$ (or lower) metallicity is extremely low and does not match with the gas-metallicity measured in this object \citep[i.e., $\sim5$\%~Z$_\odot$;][]{thuan2004,papaderos2006,izotov2006}, making this scenario unlikely.

Overall, there is not a straightforward explanation to justify all the characteristics of the point source SSC1 emission.
Having said that, the models that can simultaneously explain the MIRI/MRS \oiv/\neiii, \neiii/\neii\ and \siv/\neii\ are \citetalias{richardson2022} IMBH models (see Fig.\ref{fig:diags2}~and~\ref{fig:diags3}). 
While these grids are unable to predict the low \nev/\neii\ upper limit, we cannot completely exclude that \nev\ upper limit is underestimated since the point source spectrum is affected by residual fringing in Channel~3.

\subsection{Origin of the highly ionized extended emission}\label{sec:extended-emission}
As shown in Fig.~\ref{fig:emissionlines}, SBS~0335-052~E is characterized by extended high ionization emission lines such as \oiv~$\lambda$25.89 and \nev~$\lambda$14.3, localized mainly in the north-west region of the galaxy, covering SSCs 3-4-5-6 and S7 (all older than the point source SSC1; see Tab.~\ref{tab:compilationResults}), with a peak of high ionization emission located in the proximity of S7.
We highlight that \nev~$\lambda$14.3 has never been detected in BCDs before and SBS~0335-052~E's log(\nev/\neii)~$>0$ values are higher than any upper limits proposed for BCDs (e.g., \citealt{hao2009}), probably because of the insufficient sensitivity of ISO and Spitzer.
However, its optical counterpart (i.e., \nev~$\lambda$3426) \textit{has} been previously detected in a subgroup of BCDs, including SBS~0335-052~E (see \citealt{izotov2004,thuan2005,izotov2012,izotov2021,berg2021}), which is the lowest metallicity object in which this line has been detected so far.
\nev~$\lambda$3426 (as well as \nev~$\lambda$14.3,24.3) is usually considered a convincing proof of AGN activity (e.g., \citealt{abel2008,mignoli2013}), powered by the hard non-thermal AGN radiation, but has also been generally justified with radiative shock models in BCDs, using \citet{allen2008} models at solar and SMC (i.e., 20\%~Z$_\odot$) metallicities (e.g., \citealt{izotov2012,izotov2021}). 
However, as shown in Fig.~\ref{fig:diags1}, \ref{fig:diags2} and \ref{fig:diags3}, log(\nev/\neii)~$>0$ and enhanced log(\oiv/\neiii) cannot be explain by current state-of-the-art shock models at SBS~0335-052~E's metallicity (i.e., $\sim5$\%~Z$_\odot$). 

The difficulty in accounting for SBS~0335-052~E's extended high ionization emission was first described by \citet{herenz2017}, who showed a nebular \heii~$\lambda$4686 map from MUSE that brightened towards the north-west at the rim of a starburst driven super-shell (see also \citealt{kehrig2018}), the \heii\ shell that is out of our MIRI/MRS FOV.
In the following sections, we discuss in detail the different scenarios which can explain the sources of the extended MIR emission, drawing insights from previous studies. 
In particular, it is important to consider that \oiv\ has a similar I.P. to \heii, while \nev\ requires the presence of even harder radiation, corresponding to the extreme ultraviolet and soft X-ray range (see Tab.~\ref{tab:listEmissionLines}). 

\subsubsection{Pure Bursts of Star Formation}\label{sec:disc-stars}
\citetalias{martinez-paredes2023} and \citetalias{garofali2024} SSP models considered in this work, fail to reproduce SBS~0335-052~E \nev/\neii, \neiii/\neii, \siv/\neii\ and \oiv/\neiii\ line ratios, despite using different stellar libraries and IMFs (see Sec.~\ref{sec:models}).
This is not unexpected considering previous studies of this system. 

As mentioned previously in Sec.~\ref{sec:point-source}, \citet{kehrig2018} was only able to derive SBS~0335-052~E's \heii-ionizing budget via either single, rotating metal-free massive stars (see their Sec.~5.3.1) or a binary population with 0.05\%~Z$_\odot$ metallicity and a top-heavy IMF (modeled with BPASS; see their Sec.~5.3.2). 
However, a 0-0.05\%~Z$_\odot$ metallicity is unphysically low for SBS~0335-052~E (see also \citealt{wofford2021}), making this scenario unlikely.  
\citet{kehrig2018} discarded also Wolf-Rayet (WR) stars - massive stars ($>25-30$~M$_\odot$) with strong winds (e.g., \citealt{schaerer98}) - as the main source of ionization.
Despite WR features being observed in this galaxy (in S3 and in the south-west region of the galaxy, below S8; \citealt{papaderos2006,izotov2006,kehrig2018}), the large number needed to justify the \heii-ionization budget would overcome by a factor of $5-7$ the SSCs total mass, depending on the IMF choice \citep{kehrig2018}.
The same argument allows us to discard WRs as the main source of \oiv\ and thus \nev\ emission. 

\citet[][see their Sec~3.1]{wofford2021} proposed the presence of very massive stars (VMSs) - stars with masses above 100~M$_\odot$ \citep{vink2012} - from the blueshifted \ov~$\lambda$1371 detection in HST COS data, covering a 2.5" circular region centered on the peak of the UV emission (see Fig.~\ref{fig:sbs-f140lp}).
Thus, one might suspect that VMSs might represent the definite answer, hardening the radiation field and enabling highly ionized extended emission. 
However, VMSs, while hot and luminous, are known to have dense winds which become optically thick at the short wavelengths needed to power \heii\ and \nev\ ionizing photons, as demonstrated by \citet[][see also Hawcroft et al. 2025, submitted]{sander2022}. Although, a mechanism for the enhanced escape of the hard ionizing radiation from these winds has recently been proposed by \citet{roy2025}, whereby the winds of massive stars are inhomogeneous due to intrinsic instabilities of the outflow. \citet{roy2025} propose that the higher the degree of inhomogeneity in the outflow, the higher the emergence of extreme UV fluxes.
While this possibility has not yet been comprehensively evaluated, it may motivate revisions of stellar population models in the future.

Finally, a completely different scenario to explain the high ionization extended emission could be older ionizing stars, as suggested by previous works (e.g., \citealt{izotov1997,izotov2006}). \citetalias{martinez-paredes2023} low-metallicity SSPs models with ages of $t\geq100$~Myr (upper limit on age of SBS~0335-052~E; \citealt{izotov1997,papaderos1998}), would be dominated by post-AGB stars (specifically HOLMES) that could be capable of ionizing the gas with their strong UV continuum \citep{stasinska2008}. Indeed, \citetalias{martinez-paredes2023} (see their Fig.~2) showed that HOLMES have the peak production rate of H-ionizing photons exactly at 100~Myr. 
Interestingly, \citetalias{martinez-paredes2023} low-metallicity 100~Myr SSP models would perfectly lie on the observed \nev/\neii, \neiii/\neii, \siv/\neii\ and \oiv/\neiii\ line ratios shown in Fig.~\ref{fig:diags1},~\ref{fig:diags2},~\ref{fig:diags3}.  
However, these models were disregarded because the maximum production of ionizing photons is $\sim10^{43}$~photons~s$^{-1}$ for a 1~M$_\odot$ stellar population at SBS~0335-052~E metallicity - implying that too large a population of HOLMES would be needed to match the ionization budget estimated by \citet[][i.e., $Q(H)\sim3.86\times10^{53}$~photons~s$^{-1}$]{kehrig2018}. 
While these stars cannot be easily observed in a galaxy as distant as SBS~0335-052~E, they are not revealed in large numbers even in nearby elliptical galaxies with a weak UV excess (e.g., \citealt{brown2008}), suggesting that they do not represent a significant source of UV emission as proposed.

Overall, we have discussed how state-of-the-art SSPs are not capable of reproducing high ionization extended emission observed in SBS~0335-052~E. We also highlight that SSPs struggle to explain lower ionization MIR line ratios, such as \siv/\neii\ versus \neiii/\neii\ (see Fig.~\ref{fig:diags2}), even for values typically observed in BCDs with Spitzer (i.e., \citealt{groves2008} relation in magenta). 
This implies either a general hindering problem in low metallicity massive stars in current stellar libraries or the real need for a further source of ionization in galaxies like SBS~0335-052~E.

\subsubsection{Ultra Luminous X-ray sources}\label{sec:disc-ulx}
A ULX is observed in SBS 0335-052~E \citep{prestwich2013}, as shown in Fig.~\ref{fig:sbs-f140lp}.
In particular, ULX sources are X-ray binaries with $L>10^{39}$~erg/s and the ULX fraction seems to anti-correlate with gas-phase metallicity (e.g., \citealt{mapelli2010,mapelli2011,prestwich2013}). 
This could suggest that, given the characteristic low-metallicity gas of BCDs, ULXs could indeed play a role in their high-ionization emission.

The contribution of these X-ray sources were accurately included for the first time in the SXP grids of \citetalias{garofali2024}, which overlap with the range of \nev/\neii\ ratios (including the values of the spectrum extracted at the ULX position, magenta diamond) observed in SBS 0335-052~E at the highest ionization parameter values (i.e., log($U$)~$\sim-1.5, -1$) and for stellar population ages between 5~Myr and 20~Myr (see Fig.~\ref{fig:diags1}). 
However, the same grids over-predict the corresponding \neiii/\neii\ line ratios and, moreover, are not capable of reproducing log(\oiv/\neiii)~$>0$ that we observe (Fig.~\ref{fig:diags3}).
We stress that \citetalias{garofali2024} state-of-the-art model prescriptions could suffer from some inherent assumptions, such as the assumed theoretical scaling relationships linking X-ray luminosity with age and metallicity, while observationally there is significant scatter (see e.g., \citetalias{garofali2024} Fig.~1).

Another factor to consider is that the ULX location determined by \citet{prestwich2013} does not correspond to either the peak of the high ionization emission that we observe in the MIR or to the peak of the \heii\ and UV emission.
Indeed, \citet{kehrig2018} (see also \citet{wofford2021}) discarded the ULX as a possible source of \heii\ ionization.
In particular, analyzing Chandra data (see also \citealt{thuan2004}) they estimated the effective ionizing power at the peak of the X-ray emission and in other two regions of the galaxy, concluding that it is not enough to explain the observed \heii\ emission. 
Accordingly, this implies that the ULX cannot explain \oiv\ as well as \nev\ emission.

\subsubsection{Shocks}\label{sec:disc-shocks}
Shocks models are generally considered an important source of ionization in MIR diagnostic diagrams and can be due to several phenomena, including cloud-cloud collisions, the expansion of \hii\ regions, outflows and/or supernovae (e.g., \citealt{groves2004}). Given the disturbed kinematic of this galaxy (Fig.~\ref{fig:kinemissionlines}; see also \citealt{izotov2006,herenz2023}), the presence of clear shells in the \ha\ and \heii\ morphology \citep{herenz2017,kehrig2018} as well as diffuse non-thermal radio emission interpreted as an ensemble of compact supernova remnants expanding in a dense ISM \citep{hunt2004}, it is plausible that shocks are present and can play an important role. 
Unfortunately, the shell structure clearly visible in the \heii\ morphology is not covered in our MIRI/MRS observations. This region is slightly north-west to the area where we see $\sim50$~km/s redshifted velocities and intrinsic velocity dispersion enhancement in \siv\ and \neiii\ (Fig.~\ref{fig:kinemissionlines}).

Having said that, as shown in Figs.~\ref{fig:diags1}, \ref{fig:diags2} and \ref{fig:diags3}, low-metallicity \citetalias{flury2024} (and also \citetalias{alarie2019}, not shown) shock models do not overlap with our data. The grids could match the range of observed \oiv/\neiii\ (and other high-ionization line-ratios, such as \nev/\neiii\ and \oiv/\hpfa), but cannot explain the corresponding \nev/\neii, \neiii/\neii\ and \siv/\neii\ line ratios. 
Overall, shock models seem to overpredict low-ionization lines such as \neii. Another hint towards shocks not being the primary source of ionization of the extended \oiv\ and \nev\ emission is the complete lack of co-spatial \feii\ emission (see Sec.~\ref{sec:point-source}).
Finally, \citet{kehrig2018} also excluded shocks as the cause of the \heii\ emission, given the lack of high $S/N$(\oi~$\lambda$6300) and lack of low \sii/\ha\ line ratios, generally considered to be good shock indicators (e.g., \citealt{kewley2019,mingozzi2024}).

There might be other factors to consider when explaining the misalignment of shock models and extended high ionization emission.
First, shock fronts have spatial scales much below our spatial resolution (down to $\sim10$~pc; \citealt{calzetti2004}), which could dilute shock-driven line ratios with other ionizing sources when integrating over spatial scales larger than a shock front.
Second, it is possible that new shock models with an updated treatment of the radiative transfer with respect to Mappings~V could be in better agreement with our data (e.g., \citealt{godard2024}). 
However, these new models have not been tested yet at low-metallicity and currently take into account only velocities below 500~km/s.
Overall, we cannot completely exclude the presence of shocks, but according to the available model predictions, they do not represent the main source of ionization of MIR lines in SBS~0335-052~E.

\subsubsection{The possibility of an IMBH}\label{sec:disc-imbh}
Recent works have suggested the presence of a $10^3-10^5$~M$_\odot$ IMBH in SBS~0335-052~E, due to the detection of \nev~$\lambda$3426 \citep{hatano2024}, as well as possible variability in NEOWISE data ($3-4$~$\mu$m) over 12 years \citep{hatano2023}. 
Interestingly, the \nev~$\lambda$3426 was first observed by \citealt{thuan2005} and does not show any sign of variability \citep{hatano2024}.
\citet{hatano2023} also proposed the presence of broad \ha\ emission possibly due to the BH broad line region, from which they estimate a $M_{BH} \sim 10^8$~M$_\odot$ upper limit. 
However, this value is much larger than the mass of the different components of this galaxy (see Tab.~\ref{tab:compilationResults}), raising doubts that this emission is tracing the BH broad line region (see also Sec.~\ref{sec:high-z-context}), unless the system is not virialized (e.g., \citealt{bertemes2025}). 
As such, the existence of \nev~$\lambda$3426 and the observed NIR variability seem the two most feasible clues from the literature to suggest the presence of an IMBH in SBS~0335-052~E.

In consideration of these suggestions of an IMBH, looking at Fig.~\ref{fig:diags1} and Fig.~\ref{fig:diags2}, the models that uniquely cover the observed \nev/\neii, \neiii/\neii\ and \siv/\neii\ line ratios that we measure in SBS~0335-052~E are the $M_{BH}=10^5$~M$_\odot$ IMBH models from \citetalias{richardson2022} with an AGN fraction 4-8\% and log($U$) between $-2$ and $-0.5$ (the closed geometry model can better reach the highest \neiii/\neii).
According to the \citetalias{richardson2022} models, we can exclude the possibility of a BH with $M_{BH}=10^3$~M$_\odot$ because this BH mass would imply too low log(\neiii/\neii) ($\leq0.5$) and log(\nev/\neii) ($\leq0$) with respect to the observed data, regardless of the other parameter selection (i.e., AGN fraction, geometry, gas density, stellar age).
We can also exclude \citetalias{richardson2022} $M_{BH}=10^4$~M$_\odot$ grids, which still cannot reach the log(\nev/\neii) values that we observe ($\leq0.3$). 
\citetalias{richardson2022} $M_{BH}=10^6$~M$_\odot$ can overlap with the observed \nev/\neii\ vs \neiii/\neii\ at 8\% AGN fraction, but at lower/higher fractions, they predict larger/lower \neiii/\neii\ than what we observe. 

A great advantage of the MIRI/MRS data is that we can spatially resolve line ratios and trace the ionization hardness (e.g., \nev/\neii, \oiv/\neiii), finding their peak around/north SSCs 4,5 and S7 and with decreasing values towards south-east (see Fig.~\ref{fig:lineratios}).
As already discussed in Sec.~\ref{sec:kinmaps}, this peak region is characterized by the most disturbed kinematics with the presence of gas flows (Fig.~\ref{fig:kinemissionlines}; see also \citealt{izotov2006,herenz2023}). 
In particular, the \siv, \neiii\ and \nev\ velocity dispersion (corrected for instrumental broadening) maps revealed an interesting ``V shape" in the north-west region, starting at the position of SSCs 4-5, high-resolution optical IFU data revealed doubled-peak \ha\ and \heii\ profiles, suggesting the presence of an expanding shell \citep{izotov2006}.
In this scenario, SSC5 may correspond to the location of the IMBH.
Another potentially relevant element is that SSC5 is also found to have a very compact ($<10$~pc) NIR emitting region not resolved in HST NICMOS (high resolution, 0.075~"/px) data \citep{thompson2009}.

However, we do stress that even the \citetalias{richardson2022} models do not perfectly overlap with our data, since they are not able to reproduce the combination of highest \oiv/\neiii\ and \neiii/\neii\ values (see Fig.~\ref{fig:diags3}). 
Many factors could be at play here with regard to the assumptions and parameters held within the models - each of which are difficult to constrain with emission-line ratio diagrams alone due to the degeneracies between them. 
For instance, \citetalias{richardson2022} predictions can vary as a function of the considered SEDs (\citetalias{richardson2022} considered two extreme values, qso or disk-plaw) and assumptions on accretion rate and spin (incorporated as fixed parameters in the qso SED), as well as the choice of geometry (open vs closed). 
Also, we stress that \oiv/\neiii\ can be affected by the wavelength-dependent MIRI/MRS PSF variation correction, as well as residual fringing (see Sec.~\ref{sec:diag-ratios}).

\subsection{Implications for high-$z$ galaxies and UV-optical diagnostics}\label{sec:high-z-context}
An unexpectedly large population of objects with possible accreting black holes has started to be revealed at high-$z$ ($z\sim6-12$) with masses up to $M\sim10^8$~M$_\odot$ \citep[e.g.,][as introduced in Sec.~\ref{sec:intro}]{furtak2023,harikane2023,labbe2023,scholtz2023,larson2023,greene2024,greene2024b,maiolino2023b,kokorev2024,chisholm2024,hayes2024,cammelli2025}. 
Not only has this led to questions concerning their rapid formation (in just $\sim$300~Myr) and their subsequent impact on galaxy evolution, but also opened up discussions on which methods can unambiguously identify them.

As previously mentioned, one of the most popular criteria to identify AGN activity so far has been via broad permitted emission lines (e.g., \ha).
From a visual inspection of VLT/MUSE observations of SBS~0335-052~E MUSE, it appears that a broad \ha\ profile exists at both the position of cluster SSC1 (i.e., the MIR point source with no clear \nev\ detection) and at the peak of  \heii, \nev, and UV emission, with FWHM up to $\sim1200$~km/s (see Fig.~\ref{fig:muse} and App.~\ref{app:optical-muse}). This implies that it may have an extended nature, not compatible with a BH broad line region. 
In particular, both the \ha\ and \oiii~$\lambda$5007 emission lines require up to 3-4 Gaussian components to be well fitted, with also \oiii~$\lambda$5007 showing a fainter broad (up to FWHM~$\sim900$~km/s) component. 
Interestingly, broad \ha\ and the need of $>2$ components to reproduce the kinematics of the ionized gas has also been noticed in other nearby BCDs (e.g., \citealt{james2009,delValle-Espinosa2024,komarova2021}) and other high-$z$ analogs (e.g., Green Peas; \citealt{amorin2024}), with interpretations ranging from turbulent mixing layers, high-density gas, and radiatively driven winds. 
Despite the origin of this broadening still not being completely understood in these systems, there could be a connection with (some of) the broad-line galaxies revealed at high-$z$.

\begin{figure}[t]
\begin{center}
    \includegraphics[width=0.5\textwidth]{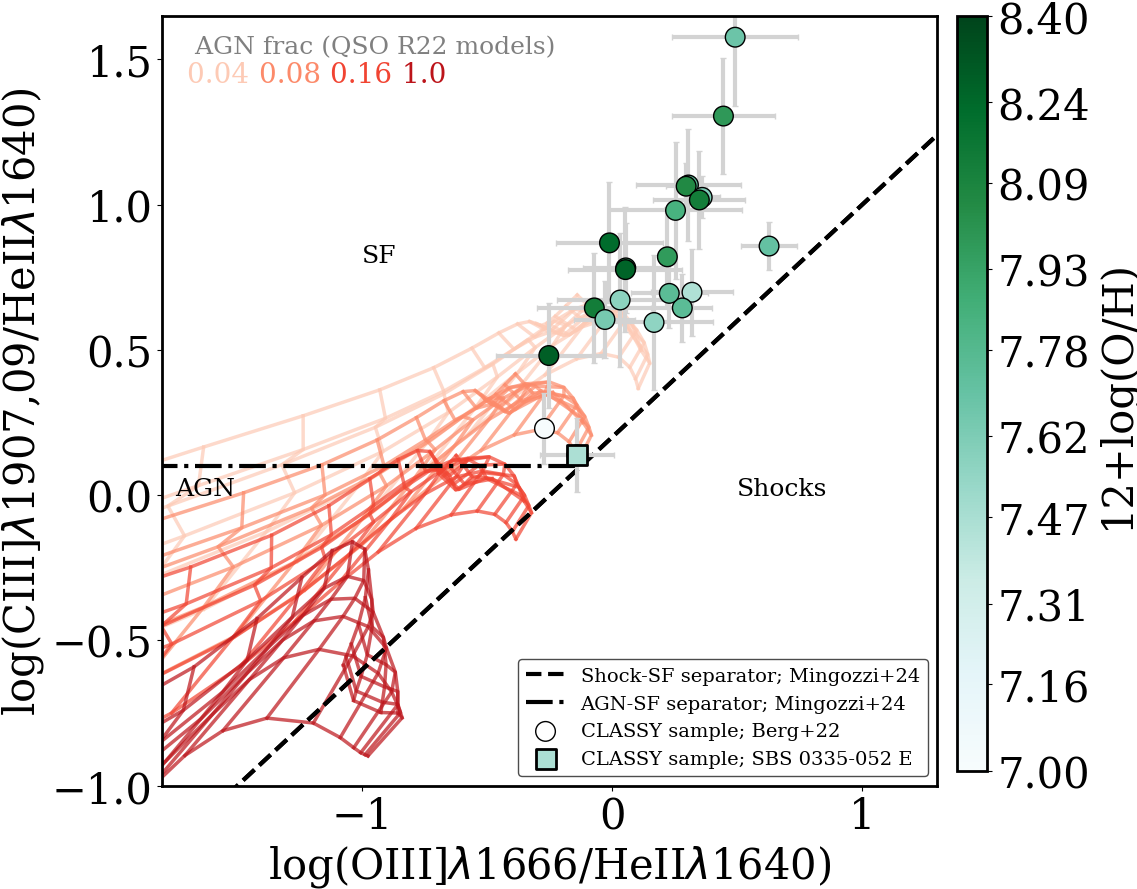}
\end{center}
\caption{\ciii~$\lambda\lambda$1907,9/\heii~$~\lambda$1640 versus \oiiiuv~$\lambda$1666/\heii~$\lambda$1640 UV diagnostic diagram, showing SBS~0335-052~E as well as other nearby high-$z$ analogs from the CLASSY survey \citep{berg2022,james2022}, color-coded as function of gas-phase metallicity. The models superimposed are the \citetalias{richardson2022} models that best-reproduced the MIR line ratios, thus enabling us to assess how the model UV line ratios behave within a metallicity range consistent with the CLASSY sample. This UV-based diagnostic diagram is the best in discriminating SF, AGN and shocks at sub-solar metallicities according to \citet{mingozzi2024}. SBS~0335-052~E UV line ratios, falling in the SF locus, are also consistent with the \citetalias{richardson2022} models with $M_{BH} = 10^5$~M$_\odot$ and 8\% AGN fraction, that reproduced the MIR line ratios. Regardless of this degeneracy, this diagram should still be regarded as an important tool in classifying the ionization source of galaxies based on their UV line ratios.} 
\label{fig:uvcomp}
\end{figure}

Exploiting the use of high-$z$ analogs, \citet{mingozzi2024} used narrow lines in optical and UV spectra from the Cosmic Origins Spectrograph (COS) Legacy Archive Spectroscopic SurveY (CLASSY; \citealt{berg2022,james2022}) to provide a toolkit to distinguish star formation and AGN activity at high-$z$. 
In particular, CLASSY is a treasury of 45 nearby ($0.002 < z < 0.182$) star-forming galaxies, including SBS~0335-052~E, generally characterized by lower stellar masses, higher SFRs and more extreme ionization fields than $z\sim0$ objects, as typically observed in the EoR.
Specifically, \citet{mingozzi2024} proposed the \ciii~$\lambda\lambda$1907,9/\heii~$\lambda$1640 versus \oiiiuv~$\lambda$1666/\heii~$\lambda$1640 as the most reliable UV diagnostic diagram (their Fig.~6; see also \citealt{feltre2016}), being capable of separating the SF, AGN and shocks at low-metallicity. 
We revisit this diagnostic diagram in Fig.~\ref{fig:uvcomp}, showing the CLASSY sample compared with \citetalias{richardson2022} models with the same parameters presented in Figure~\ref{fig:diags1} and discussed in Section~\ref{sec:disc-imbh}.
The fact that SBS~0335-052~E (light-green square) lies very close to the line separators was interpreted by \citet{mingozzi2024} as being due a combination of its low-metallicity environment and hard stellar radiation field.
However, the \citetalias{richardson2022} $M_{BH}=10^5$~M$_\odot$ IMBH model with 8\%\ AGN fraction at the lowest metallicity and highest log($U$) would also explain the UV line ratios observed in SBS~0335-052~E. 
Since the UV emission line fluxes are from spectra integrated over the COS aperture (2.5"; see Fig.~\ref{fig:sbs-f140lp}) covering the \nev~$\lambda$14.3 emitting region, the fact that the UV line ratios can be reproduced by the same set of models that explain part of the spatially resolved MIR line ratios indeed strengthens the validity of the \citetalias{richardson2022} models in describing this object.
This finding also highlights a previously unknown degeneracy of this diagnostic diagram in distinguishing between pure star-forming objects and objects with an AGN fraction $\sim4-8$~\% (when considering \citetalias{richardson2022} models with $M_{BH}=10^{3-5}$~M$_\odot$).

\begin{figure}[t]
\begin{center}
    \includegraphics[width=0.5\textwidth]{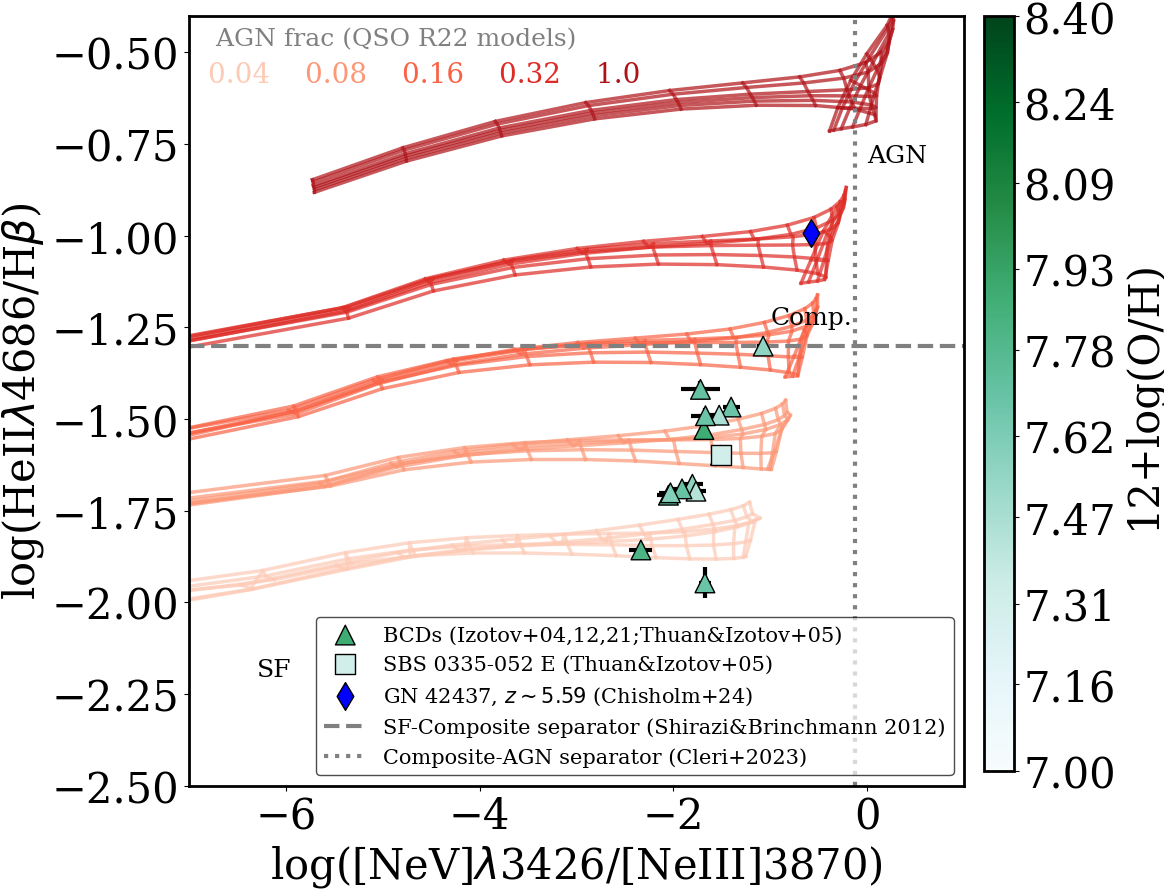}
\end{center}
\caption{\heii~$\lambda$4686/\hb\ vs \nev~$\lambda3426$/\neiii~$\lambda$3868 emission-line
diagnostic diagram, recently proposed by \citet{chisholm2024} to probe the strength of very high-ionization emission lines independently of gas-phase metallicity. The square shows SBS~0335-052~E line ratios \citep{thuan2005}, while the dots show the currently known BCDs showing optical \nev\ emission (i.e., 14, including SBS~0335-052~E; \citealt{izotov2004,thuan2005,izotov2012,izotov2021}). The blue diamond is the galaxy at $z\sim5.59$ found by \citet{chisholm2024}, to put SBS~0335-052~E and BCDs in the high-$z$ context. The dashed and dotted lines show the separators proposed by \citet{shirazi2012} and \citet{cleri2023} to distinguish SF and composite, and composite and AGN, respectively.} 
\label{fig:optcomp1}
\end{figure}

The MIR wavelength range is not observable in high-$z$ targets, but JWST does provide coverage of \nev~$\lambda3426$ in the optical wavelength range (e.g., \citealt{cleri2023,scholtz2023,chisholm2024}).
As mentioned at the beginning of Sec.~\ref{sec:discussion}, optical \nev\ has been revealed in a few nearby BCDs, including SBS~0335-052~E. Thus, it is interesting to see how SBS~0335-052~E compares to other BCDs and high-$z$ targets within which this emission has been observed.
In Fig.~\ref{fig:optcomp1} we show the diagnostic diagram \heii~$\lambda$4686/\hb\ vs \nev~$\lambda3426$/\neiii~$\lambda$3868 proposed by \citet{chisholm2024} to identify the presence of accreting IMBHs. 
Indeed, similarly to the MIR line ratios explored in Fig.~\ref{fig:diags1}, both \nev/\neiii\ and \heii/\hb\ are relatively insensitive to metallicity and instead trace the ionization parameter and the radiation hardness \citep{shirazi2012,cleri2023,chisholm2024}.
Consistently with the MIR (Fig.~\ref{fig:diags1}) and UV line ratios (Fig.~\ref{fig:uvcomp}), SBS~0335-052~E's optical line ratios are also consistent with the \citetalias{richardson2022} $M_{BH}=10^5$~M$_\odot$ IMBH models with 8\% AGN fraction.
Interestingly, SBS~0335-052~E's \nev/\neiii\ flux ratio (light-green square) is below the median values found for the 13 currently known nearby BCD \nev\ emitters \citep{izotov2004,izotov2012,izotov2021,thuan2005} and is $\sim 1/9$ of the value found for the $z\sim5.59$ galaxy studied by \citet{chisholm2024}.

Upon reflection of our comparison with UV and optical emission line diagnostics, our analysis suggests that the MIR high-ionization emission that we observe in SBS~0335-052~E could indeed be powered by an accreting IMBH. 
According to the best-fitting \citetalias{richardson2022} models, the mass of the IMBH would be $M_{BH}=10^5$~M$_\odot$, placing it slightly above the scaling relation between $M_{BH}$ and the host galaxy stellar mass, as seen for many objects at high-$z$ (e.g., \citealt{maiolino2023b}).
However, given the fact that there are several open questions regarding the predicted high-ionization emission of other models, alternative sources of ionization cannot be ruled out. In particular, it is not clear if we have completely understood the atmosphere of low metallicity massive stars (see Sec.~\ref{sec:disc-stars}), or how to constrain the different parameters within X-ray binaries, shocks and IMBHs models (see Sec.~\ref{sec:disc-ulx}, Sec.~\ref{sec:disc-shocks}, Sec.~\ref{sec:disc-imbh}). As such, our model-based conclusions can only be as firm as the models are realistic. 

\section{Conclusions}\label{sec:conclusion}
In this paper, we reported the first JWST MIRI/MRS data of a BCD, focusing on the galaxy SBS~0335-052~E. 
This object is well-known for its extremely metal-poor environment, complex kinematics, high-ionization optical and UV emission lines, and the possible presence of an IMBH, which makes it the perfect analog of high-$z$ systems in the EoR.
In particular, we used MIRI/MRS data to spatially resolve the MIR emission in this BCD (down to $\sim0.4−0.9$", $\sim112−252$~pc) through emission lines tracing different levels of ionization (e.g., \neii, \siv, \neiii, \oiv, \nev) of the ionized gas. In this way, we can reveal the ionization structure of this galaxy and investigate the possible accreting IMBH or other non-stellar ionizing sources, building on the many previous works carried out since SBS~0335-052~E discovery \citep{izotov1990}.
In the following, we summarize our main findings:
\begin{itemize}
    \item We revealed the presence of a point-source-like emission dominating the continuum and emission-line fluxes close to the position of one of the youngest and most embedded stellar clusters, SSC1 ($t\sim3$~Myr, $A_V\sim15$; see Fig.~\ref{fig:sbs-f140lp}). We accurately modeled and isolated its emission (see Fig.~\ref{fig:1dspecpsf}; Sec.~\ref{app:psf-sub}), to study in detail its properties as well as the underlying extended emission. 
    The point-source spectrum does not show very high-ionization emission (i.e., \nev~$\lambda$14.32), and is characterized by the strongest silicate feature at 9.7~$\mu$m, H$_2$ molecular gas emission and low ionization lines (i.e., \feii~$\lambda$5.34). As we discuss in Sec.~\ref{sec:point-source}, there is not a straightforward explanation to explain all its characteristics. \citetalias{richardson2022} $M_{BH}=10^5$~M$_\odot$ IMBH models can well reproduce the detected MIR lines, but cannot explain the low \nev/\neii\ upper limit.
    \item We revealed extended high ionization emission lines such as \oiv~$\lambda$25.89 and \nev~$\lambda$14.32, localized mainly in the north-west region of the galaxy, covering SSCs 3-4-5-6 and S7 (all older than the point source SSC1). In particular, this is the first \nev~$\lambda$14.32 detection in a BCD (Sec.~\ref{sec:line-maps}, Fig.~\ref{fig:emissionlines}). 
    \item We spatially resolved line ratios tracing the ionization hardness (e.g., \nev/\neii, \oiv/\neiii), finding their peak around/north SSCs 4,5 and S7 and with decreasing values towards south-east (Sec.~\ref{sec:line-ratios}, Fig.~\ref{fig:lineratios}). Interestingly, we found log(\nev/\neii)~$>0$, higher than any upper limit proposed for BCDs with previous IR telescopes (Sec.~\ref{sec:line-ratios}, Fig.~\ref{fig:lineratios}).
    \item We showed how the region at the peak of the ionization hardness is characterized by the most disturbed kinematics with the presence of gas flows and enhanced velocity dispersion, consistent with the presence of an expanding shell as suggested in previous works (Sec.~\ref{sec:kinmaps}; Fig.~\ref{fig:kinemissionlines}). 
    \item We explored MIR diagnostic diagrams, comparing MIR line ratios with a set of state-of-the-art photoionization and shock models, finding that SBS~0335-052~E lies outside of the “canonical" SF locus and can be best reproduced by \citetalias{richardson2022} $n_H=100$~cm$^{-3}$ $M_{BH}=10^5$~M$_\odot$ IMBH models with AGN fraction $<16$\% (Sec.~\ref{sec:diag-ratios}; Fig.~\ref{fig:diags1}, Fig.~\ref{fig:diags2}). However, these models struggle to reproduce the highest \oiv/\neiii\ MIR line ratios (Fig.~\ref{fig:diags3}). 
    \item The \citetalias{richardson2022} models are consistent also with UV and optical emission-line diagnostics (Sec.~\ref{sec:high-z-context}; Fig.~\ref{fig:uvcomp}, Fig.~\ref{fig:optcomp1}), strengthening the possibility of the presence of an accreting IMBH. 
    On the other hand, star-forming models (regardless of including X-ray binaries) and shocks struggle to reproduce even MIR line ratios tracing lower ionization than \oiv\ and \nev\ (Fig.~\ref{fig:diags2}), typically observed in BCDs, suggesting possible model limitations.
    In Sec.~Sec.~\ref{sec:extended-emission} (see also \ref{sec:high-z-context}) we also highlight how there are several open questions regarding the predicted high-ionization emission of other models, implying that alternative sources of ionization (e.g., very massive stars, Sec.~\ref{sec:disc-stars}; Ultra Luminous X-ray sources, Sec.~\ref{sec:disc-ulx}; shocks, Sec.~\ref{sec:disc-shocks}) cannot be completely ruled out. 
\end{itemize}
Overall, until models are adapted to fully replicate the ionizing sources, one way to step-forward in our understanding of the role that objects such as SBS~0335-052~E have in interpreting the high-$z$ universe would be to perform detailed multi-wavelength modelling, i.e. utilizing the full X-ray--sub-mm information as constraints. Finding more near and far \nev\ emitters, as well as investigating more dwarf AGN candidates (e.g., \citealt{polimera2022,mezcua2024,wasleske2024}), covering a diverse range of properties, will also help to better understand the conditions required to produce this high-ionization emission.

\acknowledgments
This work is based on observations made with the NASA/ESA/CSA James Webb Space Telescope. The data were obtained from the Mikulski Archive for Space Telescopes at the Space Telescope Science Institute, which is operated by the Association of Universities for Research in Astronomy, Inc., under NASA contract NAS 5-03127 for JWST. These observations are associated with program JWST-4278. The {\it JWST/MIRI-MRS} data used in this paper can be found in MAST: \dataset[https://doi:10.17909/v4tb-5y16]{https://doi:10.17909/v4tb-5y16}.
MM, BLJ and SH are thankful for support from the European Space Agency (ESA).
MJH is supported by the Swedish Research Council (Vetenskapsr{\aa}det), and is fellow of the Knut \& Alice Wallenberg Foundation. 
RA acknowledges the support of project PID2023-147386NB-I00 and the Severo Ochoa grant CEX2021-001131-S funded by MCIN/AEI/10.13039/50110001103. 
This research has used the HSLA database, developed and maintained at STScI, Baltimore, USA.
MM is grateful to Carlo Cannarozzo, Giovanni Cresci, Andy Fox, and Travis Fischer for inspiring conversations and advice.

\facilities{JWST (MIRI), HST (COS, ACS), VLT (MUSE)}
\software{
astropy (The Astropy Collaboration 2013, 2018)
dustmaps (Green 2018),
jupyter (Kluyver 2016),
LINMIX (Kelly 2007) 
Photutils (Bradley 2021),
python,
pysynphot (STScI Development Team)}

\typeout{} 
\bibliography{sbs_bib}

%




\appendix
\section{Point source subtraction procedure}\label{app:psf-sub}
We modelled the PSF for each MIRI/MRS band using v~1.4.0 WebbPSF \citep{perrin2014} and the function \textit{calc\_datacube}\footnote{\url{https://webbpsf.readthedocs.io/en/latest/jwst_ifu_datacubes.html}}, taking into account the same native pixel scale and spectral binning of the observed data. 
We created the pipeline Stage 3 products by band, selecting the option ifualign, to have them aligned with the WebbPSF models. 
Then, we extracted $1~\mu$m ($1.5~\mu$m for Channel 4) slices around the emission lines taken into account in this paper (see Sec.~\ref{sec:fitting}) from the Stage 3 band datacubes. 
We localized the point source position in each slice, modeling a 2D Gaussian distribution in a region of the spectrum with solely continuum emission, where the point-source structure is more enhanced (emission lines have also extended emission). 
Then, using photutils PSFphotometry we found the best-fit model of the point-source emission in each slice, wavelength-by-wavelength, putting very tight constraints on its position and allowing the PSF model normalization to vary to match the flux in a region $\sim 2$ times the PSF FWHM. According to WebbPSF and JWST documentation, the PSF FWHM in Channel 1, 2, 3 and 4 is $\sim 0.4", 0.5", 0.6", 0.9"$ (i.e., 2, 2.5, 3, 4.5~px), respectively. 

\begin{figure}[!h]
\begin{center}
        \begin{minipage}{0.45\textwidth}
            \centering
                \includegraphics[width=1\textwidth]{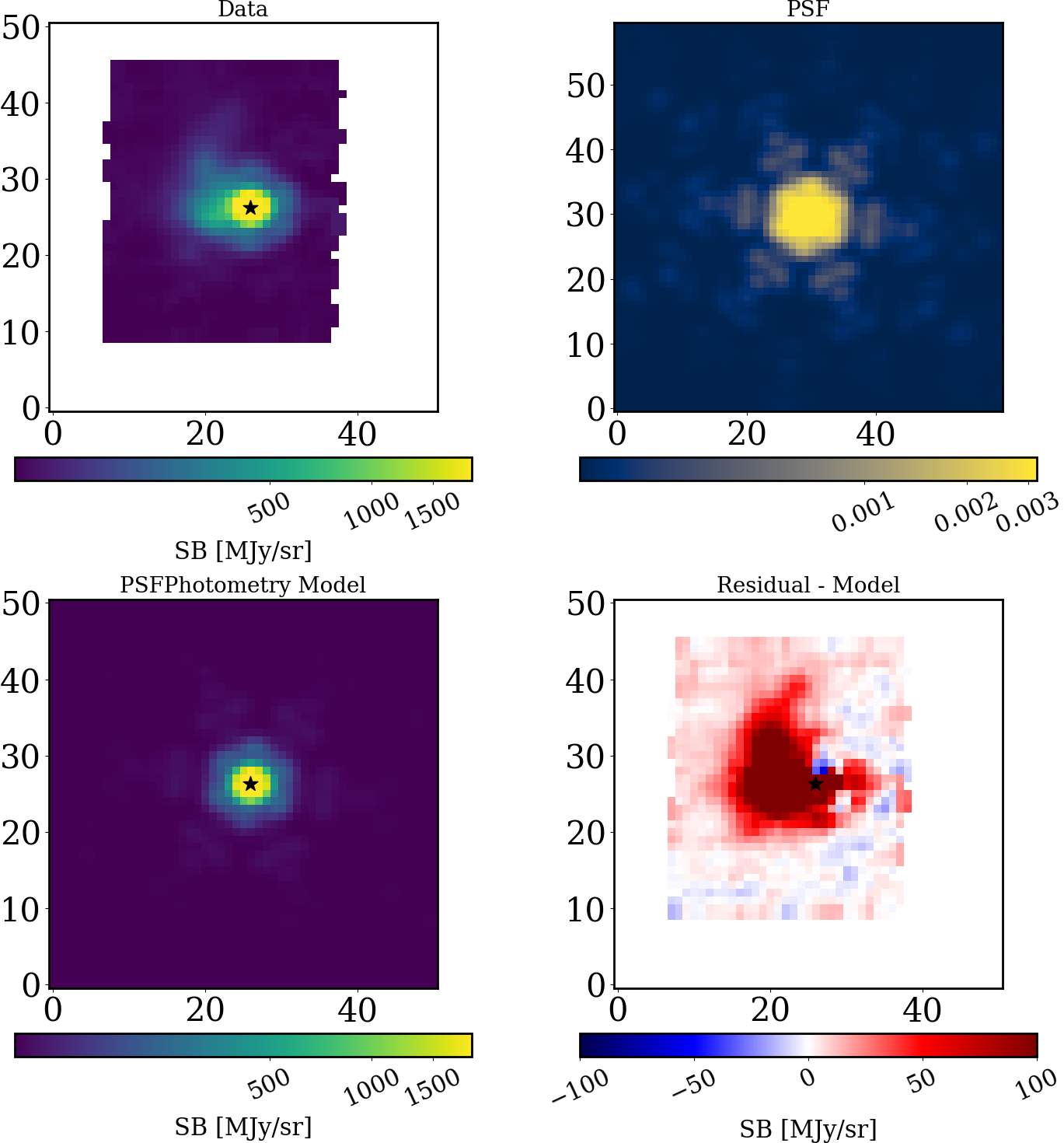}
        \end{minipage}
        \hfill
        \begin{minipage}{0.45\textwidth}
            \includegraphics[width=.9\textwidth]{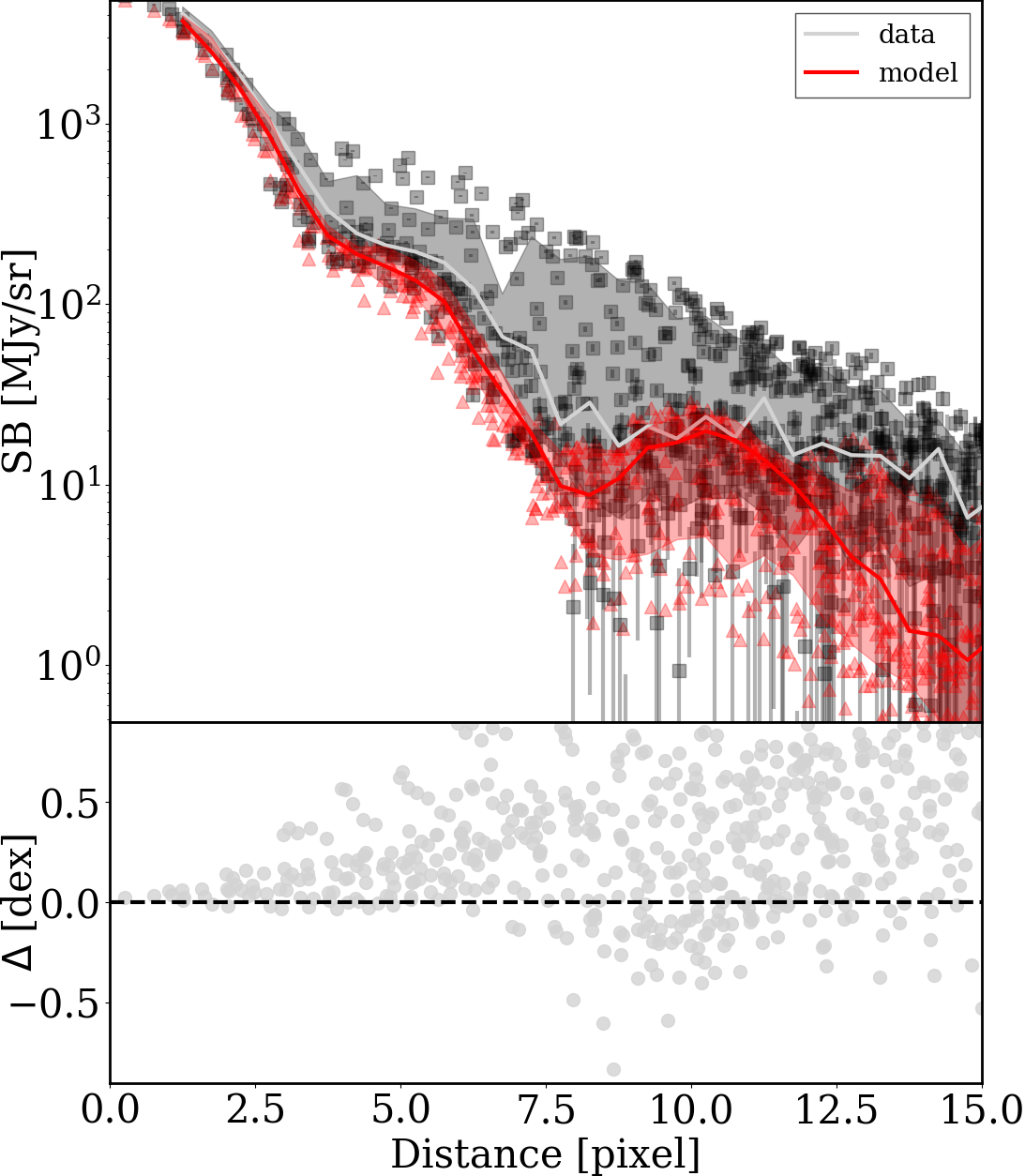}
            
        \end{minipage}
\end{center}
\caption{Left panels: images from one of the original Channel 3C slices capturing part of the \neiii\ line point-like and extended emission, with corresponding WebbPSF model, best-fit model and final residuals. The black star indicates the position where the PSF model is centred and then subtracted. The images have the orientation ``ifualign" and north is to the left. Right panel: radial distribution of the observed surface brightness SB (black), best-fit model (in red) and residuals (gray dots, bottom panel) in the Channel 3C slice shown above. 
The lack of PSF structure in the residual map and the good match between the radial distributions within $\sim3$~px from the point source (Channel 3C PSF FWHM $\sim 0.6" \sim 3$~px) validate our point-source subtraction procedure.}
\label{fig:psf-sub}
\end{figure}

Fig~\ref{fig:psf-sub} displays an example of the data, the original WebbPSF model, the best-fit model we find with photutils and the residual extended emission for one wavelength close to the \neiii~$\lambda$15.56 emission, where both point-like source and extended emission are clearly visible (see upper left panel). 
The lack of PSF structure in the residual map and the good match between the radial distributions within $\sim3$~px from the point source location (Channel 3C PSF FWHM $\sim 0.6" \sim 3$~px) validate our point-source subtraction procedure. 

\begin{figure*}
\begin{center}
    \includegraphics[width=0.8\textwidth]{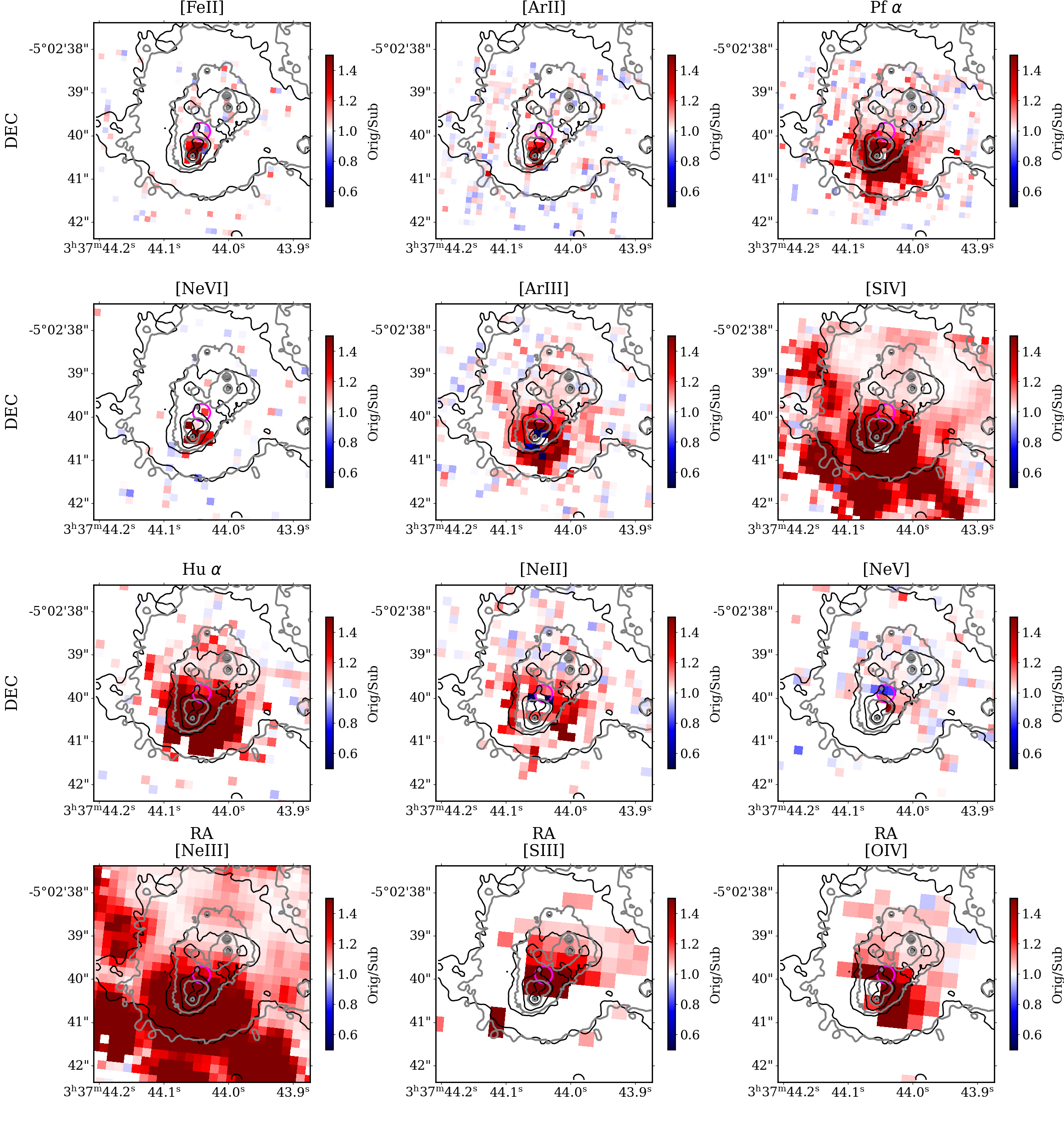}
\end{center}
\caption{Ratio of the original and point-source subtracted maps, highlighting the region dominated by the point source (darker red). All the spaxels with $S/N>3$ are shown. \neiii\ and \siv\ are the brightest emission lines observed and show clearly the MIRI/MRS PSF structure. The black and gray contours show the H$\alpha$ and UV emission, while the magenta circle shows the position of the ULX identified in this galaxy, as shown in Fig.~\ref{fig:sbs-f140lp}. North is up, east is to the left. The emission-line maps are shown at their native pixel scale, without PSF convolution applied.}
\label{fig:psf-maps}
\end{figure*}

Fig.~\ref{fig:psf-maps} shows the ratio between the original and the PSF subtracted emission-line maps, to highlight which regions and emission lines are mainly affected by the contribution of the point source (darker red). The fitted emission lines are shown in order of their wavelength (see Tab.~\ref{tab:listEmissionLines}).
Overall, the point source emission seems to be located in SSC1, as we discuss in Sec.~\ref{sec:point-source} (see also Fig.~\ref{fig:sbs-f140lp}).

We applied the same point-source subtraction procedure on the nine band datacubes (A, B, C) of Channel 1, 2 and 3, to retrieve the point-source subtracted spectra for the extended emission continuum and the point source continuum. We used these data to model the continuum with CAFE and estimate the dust attenuation, as explained in Sec.~\ref{sec:dust}.

\section{Other MIR diagnostic diagrams}\label{app:other-diags}
\begin{figure*}
\begin{center}
    \includegraphics[width=1\textwidth]{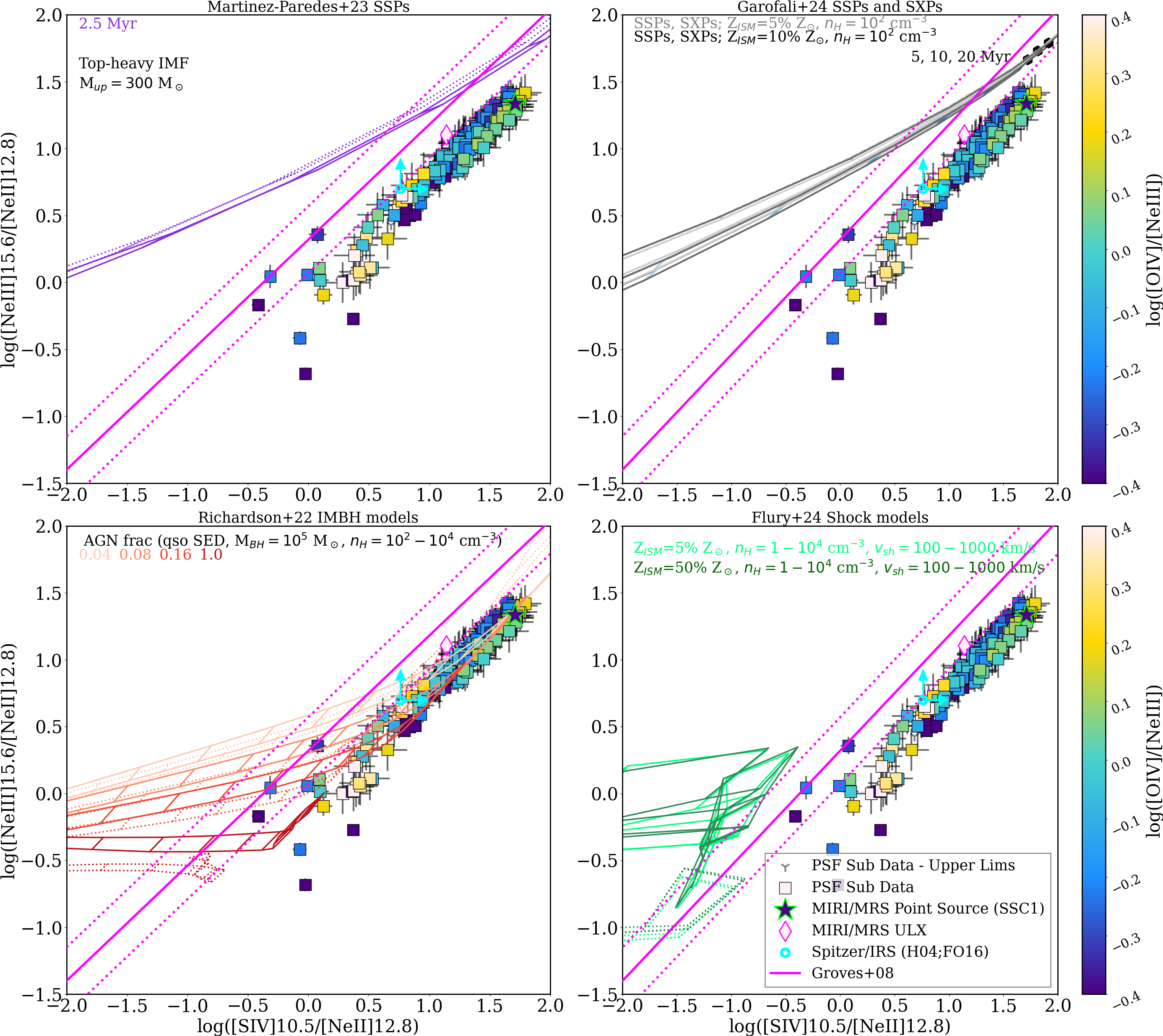}
\end{center}
\caption{\neiii/\neii\ vs \siv/\neii\ diagnostic diagram color-coded as a function of \oiv/\neiii\ line ratio, with overplotted the same four sets of models of Fig.~\ref{fig:diags1} (higher density models are shown as dotted lines). The color-coding is the same as Fig.~\ref{fig:lineratios}, to visually understand to which regions the line ratios correspond. This diagram allows us to compare with models also the regions of SBS~0335-052~E with no \nev\ or \oiv\ emission, that are not shown in Fig.~\ref{fig:diags1} and Fig.~\ref{fig:diags3}, as well as typical galaxies, including BCDs, found to follow the magenta line with some scatter \citep{groves2008}. Clearly, all the low-metallicity models apart from \citetalias{richardson2022} IMBH (with $n_H=100$~cm$^{-3}$) struggle to reproduce SBS~0335-052~E's line ratios. 
Our dust attenuation correction made log(\siv/\neii) line ratio $\sim0.1$~dex higher than the observed value.}
\label{fig:diags2}
\end{figure*}

\begin{figure*}
\begin{center}
    \includegraphics[width=1\textwidth]{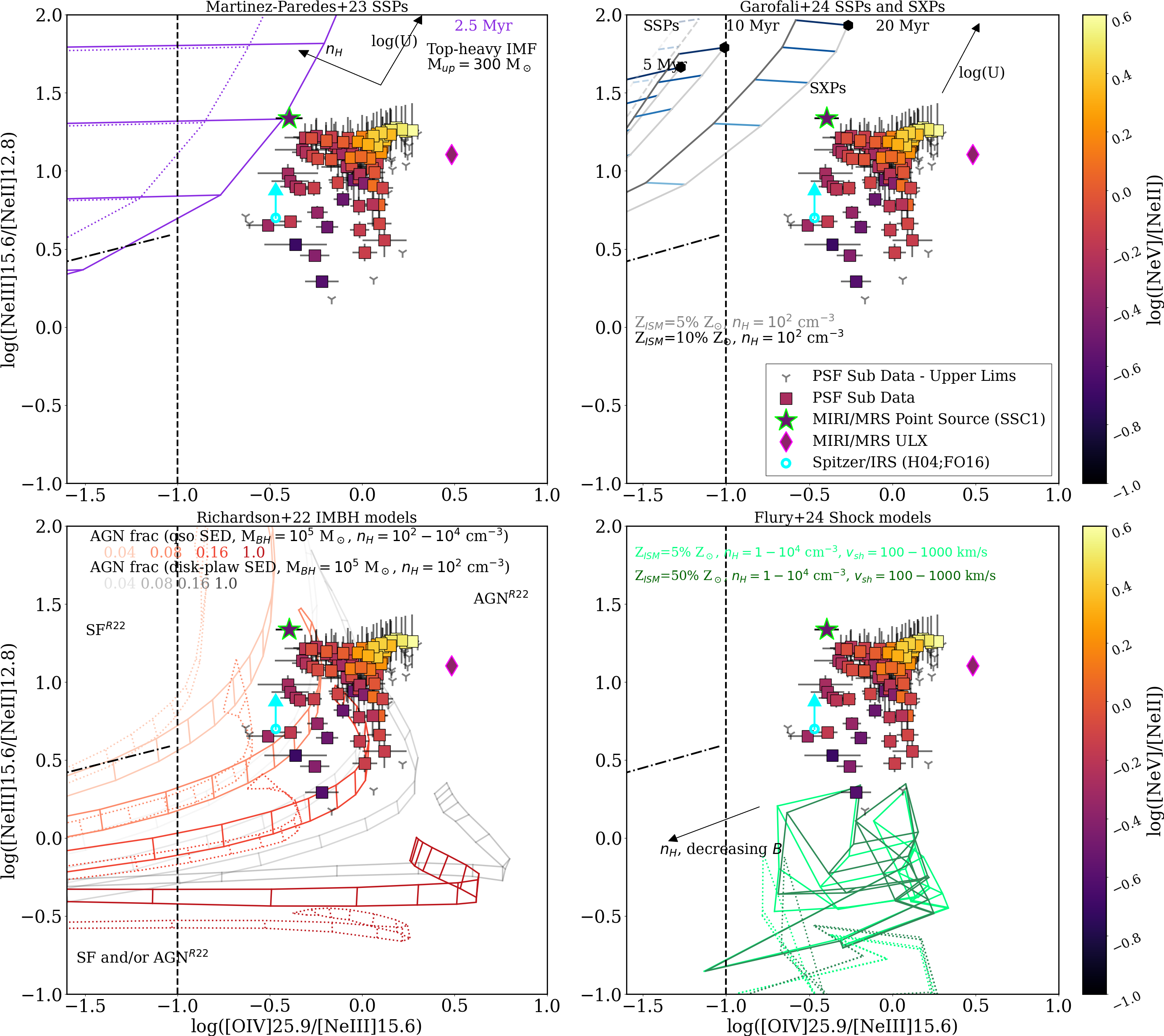}
\end{center}
\caption{\neiii/\neii\ versus \oiv/\neiii\ diagnostic diagram color-coded as a function of \nev/\neii\ line ratio, with overplotted the same four sets of models of Fig.~\ref{fig:diags1}. The separators are taken from \citetalias{richardson2022}. The color-coding is the same as Fig.~\ref{fig:lineratios}, to visually understand to which regions the line ratios correspond. \citetalias{richardson2022} IMBH low-density models with a low AGN fraction ($<16$\%) partially overlap with the data, but are not able to explain the highest \oiv/\neiii\ ratios (the disk-plaw SED can reproduce higher \oiv/\neiii\ than the qso SED). Young massive stars could explain the point-source (star symbol) line ratios, but struggle to reproduce \nev/\neii\ upper limit. This diagnostic diagram can be affected by the wavelength-dependent MIRI/MRS PSF and residual fringing, given that \neiii\ lies in Channel 3C and \oiv\ in Channel 4C.}
\label{fig:diags3}
\end{figure*}

In this section we show the \neiii/\neii\ vs \siv/\neii, color-coded as a function of \oiv/\neiii\ (Fig.~\ref{fig:diags2}) and \neiii/\neii\ versus \oiv/\neiii, color-coded as a function of \nev/\neii\ (Fig.~\ref{fig:diags3}), discussed in Sec.~\ref{sec:diag-ratios}. 
In the figures, we show the same grids shown in Fig.~\ref{fig:diags1}, apart from in Fig.~\ref{fig:diags3} where we also show the \citetalias{richardson2022} disk-plaw models (gray scale) that can reach slightly higher \oiv/\neiii.
We comment in detail these figures in Sec.~\ref{sec:diag-ratios} and Sec.~\ref{sec:discussion}.

We highlight that Fig.~\ref{fig:diags2} diagnostic diagram can be affected by dust attenuation since the \siv\ line lies in the deepest part of the silicate feature (highlighted in gray in Fig.~\ref{fig:1dspecpsf}, Fig.~\ref{fig:1dspeculx} and Fig.~\ref{fig:1dspecnev}). 
We applied a dust reddening correction (Sec.~\ref{sec:dust}) that makes the log(\siv/\neii) line ratio $\sim0.1$~dex higher.
Also, this diagram is unable to well-separate the low/intermediate and high-ionization emission, given that \siv\ and \neiii\ have similar I.P. and behave similarly.
However, it allows us to compare with models also the regions of SBS~0335-052~E with no \nev\ or \oiv\ emission, that are not shown in Fig.~\ref{fig:diags1} and Fig.~\ref{fig:diags3}.
The magenta line shown in Fig.~\ref{fig:diags2} reports the empirical relation found by \citet{groves2008}, valid - with some scatter - for very different galaxies, including starbursts, ultra-luminous infrared galaxies, AGN, and also BCDs.
Interestingly, SSP and SXPs models seem able to replicate the magenta line only at the highest \siv/\neii, while shocks can only account for the lowest \siv/\neii\ and \neiii/\neii. 
This means that according to the explored set of models, many BCDs could be reproduced only with IMBH models, as shown for SBS~0335-052~E.

Fig.~\ref{fig:diags3} diagnostic diagram can be more affected by the wavelength-dependent MIRI/MRS PSF and residual fringing than Fig.~\ref{fig:diags1} and Fig.~\ref{fig:diags2}, given that \neiii\ lies in Channel 3C and \oiv\ in Channel 4C. 
The effect of the PSF is also probably affecting the ULX \oiv/\neiii\ (magenta diamond), obtained by fitting the ULX spectrum extracted with a variable (PSF-dependent) aperture (see Sec.~\ref{sec:fitting}). Indeed, the ULX \oiv/\neiii\ looks $\sim0.5$~dex higher than the median value of the spatially resolved data.
\citetalias{richardson2022} IMBH low-density models with a low AGN fraction ($<16$\%) and partially overlap with the data, but are not able to explain the highest \oiv/\neiii\ ratios. 
Young massive stars may be able to explain the point-source (star symbol) line ratios but struggle to reproduce its \nev/\neii\ upper limit as well as its \siv/\neii\ vs \neiii/\neii\ line ratios shown in Fig.~\ref{fig:diags2}.

\section{The optical counterparts}\label{app:optical-muse}
\begin{figure*}[t]
\begin{center}
    \includegraphics[width=1\textwidth]{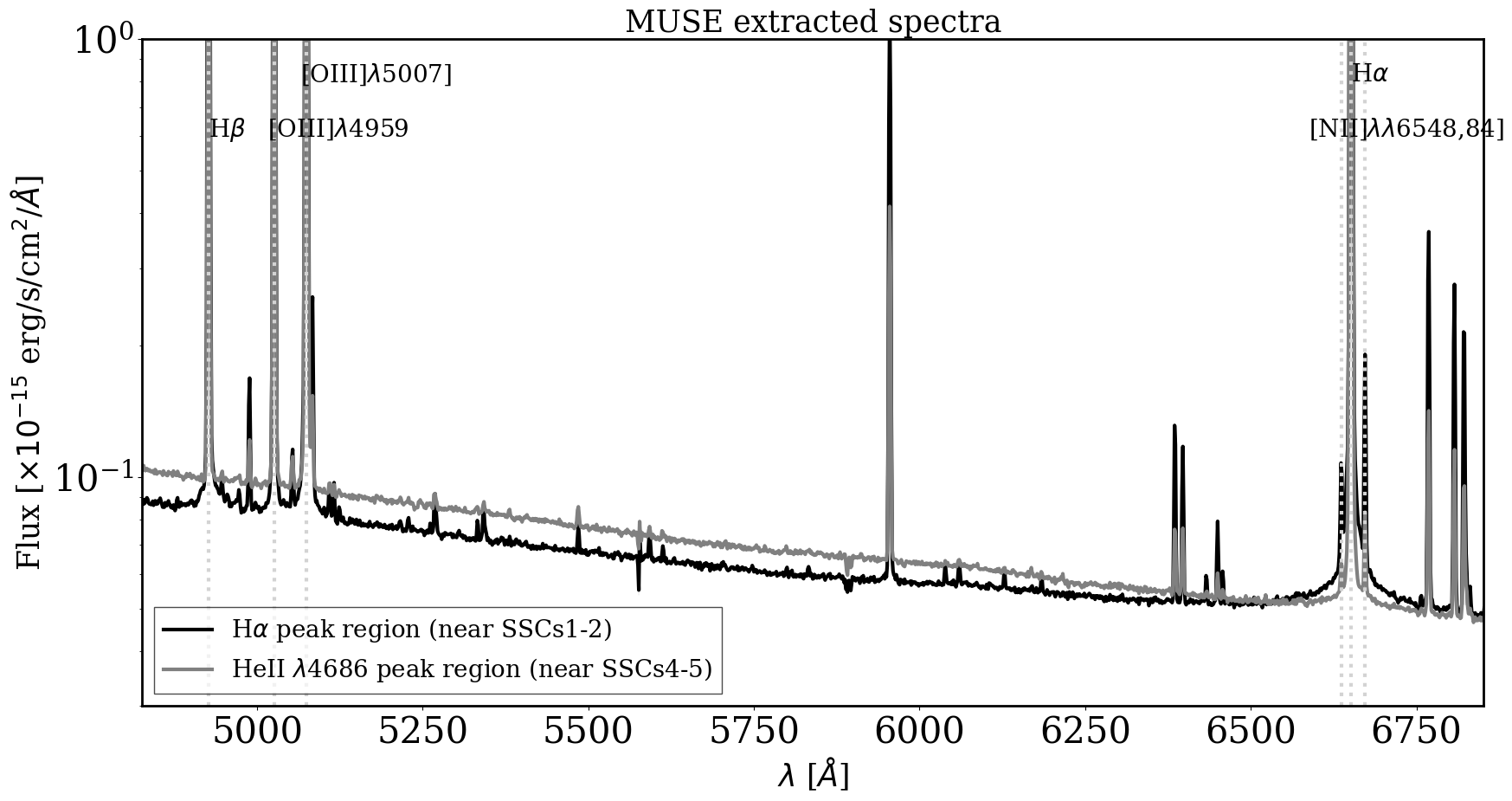}
    \includegraphics[width=0.47\textwidth]{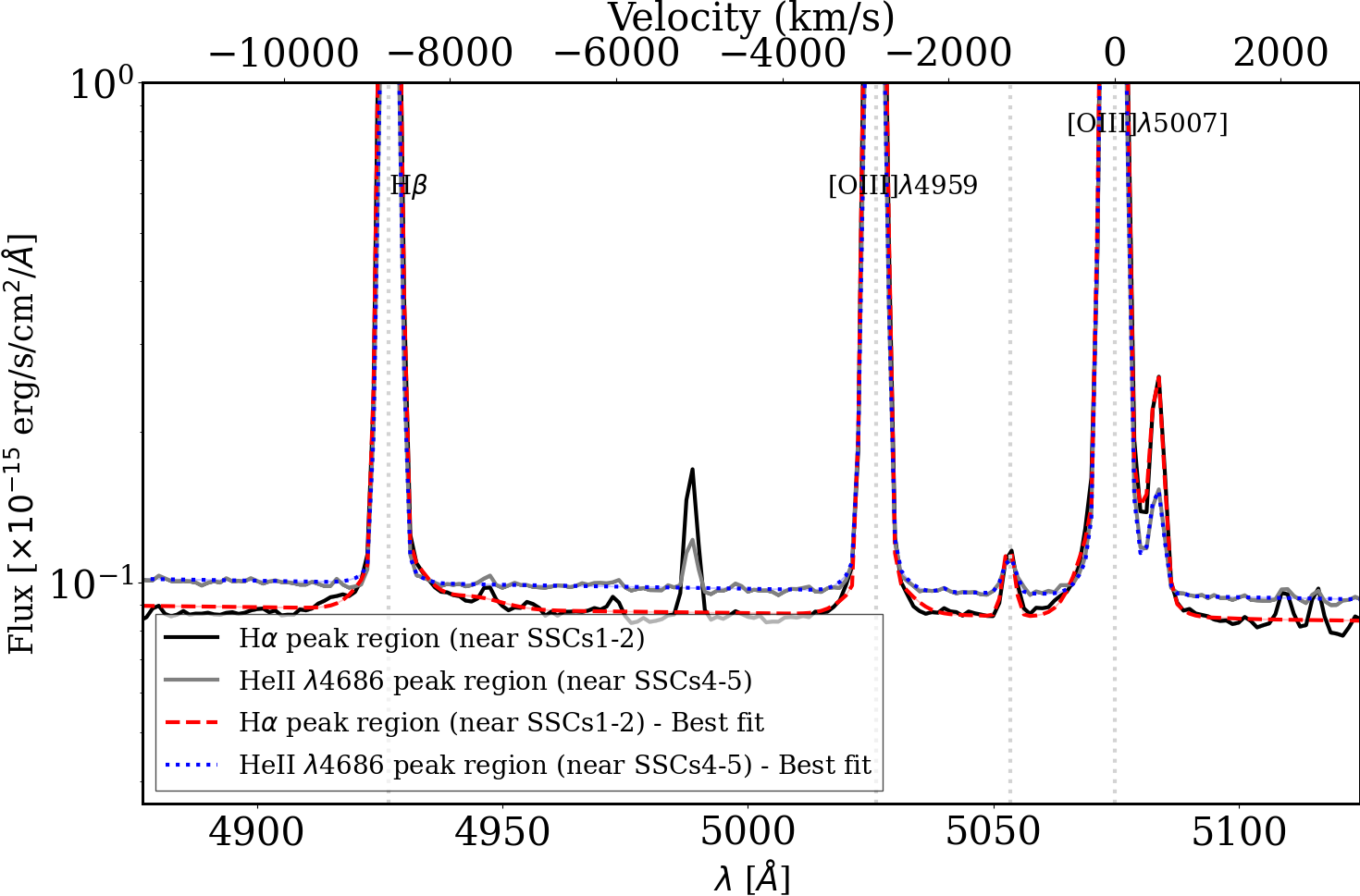}
    \includegraphics[width=0.47\textwidth]{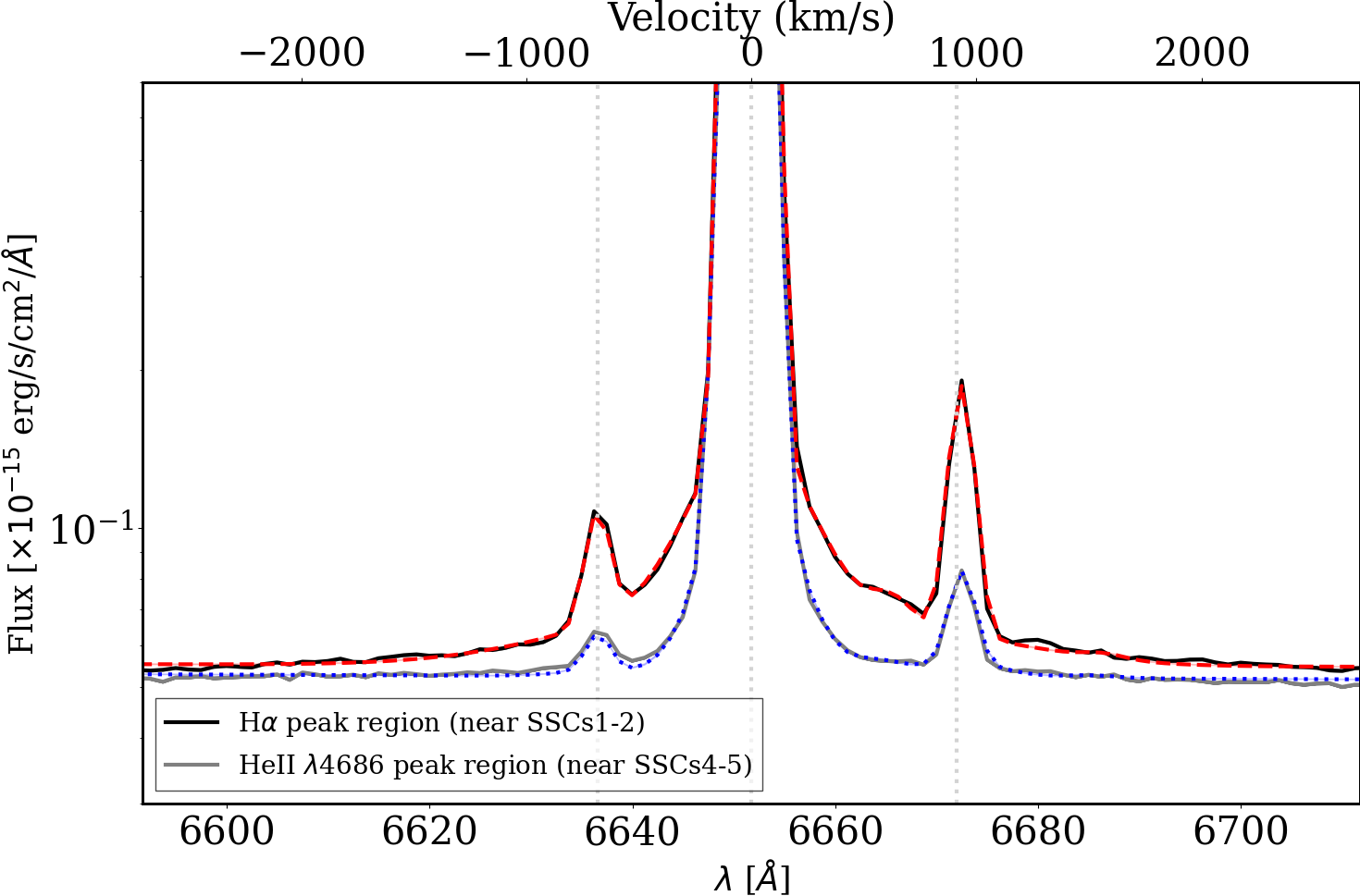}
\end{center}
\caption{Upper panel: Extracted MUSE spectra from the point-source region (SSC1; black) and the UV and \nev\ emitting region (around SSC4+5; gray), with a 3-px extraction radius. Bottom panels: Zoom on the \hb+\oiii~$\lambda\lambda$4959,5007 (left) and \ha+\nii~$\lambda\lambda$6548,84 (right). Both spectra show a broad \ha\ line that is more visible in the black spectrum, which shows also a (fainter) broad component under \hb\ and the forbidden \oiii~$\lambda$5007 emission lines, as reported in App.~\ref{app:optical-muse}. This indicates complex kinematics, but does not imply the presence of a broad line region.}
\label{fig:muse}
\end{figure*}

Fig.~\ref{fig:muse} top panel shows the region around the MIRI/MRS point-source (near SSCs1-2; in black) and the UV and \nev\ emitting region (around SSCs4-5; in gray) spectra extracted from the MUSE data (ID~096.B-0690A; PI Hayes), with a 3~px ($\sim 0.6$") extraction radius at the peak of the \ha\ and \heii~$\lambda$4686 emission, respectively.
Consistently with our MIRI/MRS findings, the black spectrum has a larger dust attenuation than the gray one, looking at the different continuum slopes.
Interestingly, the \ha\ line shows a broad profile in both spectra. This is more evident in the black spectrum, which shows also a broad (fainter) \hb\ and \oiii~$\lambda\lambda$5007. 

Fig.~\ref{fig:muse} bottom panels show a zoom around the \hb\ and \oiii~$\lambda\lambda$4959,5007 (left) and \ha\ and \nii~$\lambda\lambda6548,84$  (right) emission lines, with our best-fit shown in red and blue.
To reproduce the black and gray spectra \ha\ profiles we added up to 4 Gaussian components, with maximum widths of $FWHM\sim1200$~km/s and $FWHM\sim400$~km/s, respectively.
To fit the forbidden \oiii~$\lambda$5007 line in the black and gray spectra we added up to 3 Gaussian components, with maximum widths of $FWHM\sim900$~km/s and $FWHM\sim500$~km/s, respectively.

The full modeling and understanding of the complex kinematics highlighted in the MUSE datacube are beyond the scope of this paper. However, the need for (even broad) multiple Gaussian components in spectra located in different parts of the galaxy and in both permitted and forbidden lines suggests that there is no BH broad line region. Similar broad components ($FWHM\sim1000$~km/s) have been reported in other metal-poor galaxies (e.g., the brightest star-forming region in CGCG 007-025, \citealt{delValle-Espinosa2024}), whose extension also prevents its association with a BH broad line region.

\end{document}